%% file: nmpc_rsm.tex
\documentclass[journal]{IEEEtran}
\usepackage{hyperref}
\usepackage{lipsum}

\ifCLASSINFOpdf
  \usepackage[pdftex]{graphicx}
\else
\fi
\usepackage{amsmath}
\usepackage{url}

\usepackage{xcolor}
\usepackage{amsmath,mathtools,amssymb}
\usepackage{scalerel}
\newcommand*{\mylatexmacrospath}{./include/}

\input{\mylatexmacrospath/MyPackages_General_EN}

\input{\mylatexmacrospath/MyPackages_TikZ_externalize}


\input{\mylatexmacrospath/MyEnvironments_Paper_EN}
\input{\mylatexmacrospath/MyColors}
\input{\mylatexmacrospath/MySymbols_GeneralMath}

\input{\mylatexmacrospath/MySymbols_ControlSystems}

\input{\mylatexmacrospath/MySymbols_ThreePhaseSystems}

\input{\mylatexmacrospath/MySymbols_GridConnection}
\input{\mylatexmacrospath/MySymbols_PowerElectronics}

\input{\mylatexmacrospath/MySymbols_ElectricalDrives}
\input{\mylatexmacrospath/MySymbols_RSM_JulianKullick}

\newif\ifcommentandrea
\commentandreatrue

\usepackage{siunitx}
\begin{document}
%
% paper title
% Titles are generally capitalized except for words such as a, an, and, as,
% at, but, by, for, in, nor, of, on, or, the, to and up, which are usually
% not capitalized unless they are the first or last word of the title.
% Linebreaks \\ can be used within to get better formatting as desired.
% Do not put math or special symbols in the title.
\title{Continuous Control Set Nonlinear Model Predictive Control of Reluctance Synchronous Machines}
%
%
% author names and IEEE memberships
% note positions of commas and nonbreaking spaces ( ~ ) LaTeX will not break
% a structure at a ~ so this keeps an author's name from being broken across
% two lines.
% use \thanks{} to gain access to the first footnote area
% a separate \thanks must be used for each paragraph as LaTeX2e's \thanks
% was not built to handle multiple paragraphs
%

\author{
\thanks{This research was supported by the German Federal 
Ministry for Economic Affairs and Energy (BMWi) via 
eco4wind (0324125B) and DyConPV (0324166B), by 
DFG via Research Unit FOR 2401 and by the EU via ITN-AWESCO
(642 682)}
Andrea~Zanelli\thanks{Andrea Zanelli is with the Systems Control and Optimization Laboratory,
Department of Microsystems Engineering, University of Freiburg, Freiburg, Germany (email: andrea.zanelli@imtek.de).},
Julian Kullick\thanks{Julian Kullick 
is with the Department of Electrical Engineering and
Information Technology at the Munich University of Applied Sciences, Munich, Germany (email: julian.kullick@hm.edu).}, 
Hisham Eldeeb\thanks{Hisham Eldeeb is with IAV GmbH, Munich, Germany (email: hisham.eldeeb@iav.de).},
Gianluca Frison\thanks{Gianluca Frison is with the Systems Control and Optimization Laboratory,
Department of Microsystems Engineering, University of Freiburg, Freiburg, Germany (email: gianluca.frison@imtek.de).},
Christoph Hackl\thanks{Christoph M. Hackl is with the Department of Electrical Engineering and
Information Technology at the Munich University of Applied Sciences, Munich, Germany (email: christoph.hackl@hm.edu).},
Moritz Diehl\thanks{Moritz Diehl is with the Systems Control and Optimization Laboratory,
Department of Microsystems Engineering and Department of Mathematics,
University of Freiburg, Freiburg, Germany
(email: moritz.diehl@imtek.uni-freiburg.de).}}
% \thanks{
% of Electrical and Computer Engineering, Georgia Institute of Technology, Atlanta,
% GA, 30332 USA e-mail: (see http://www.michaelshell.org/contact.html).}% <-this % stops a space
% \thanks{J. Doe and J. Doe are with Anonymous University.}% <-this % stops a space
% \thanks{Manuscript received April 19, 2005; revised August 26, 2015.}}

% note the % following the last \IEEEmembership and also \thanks - 
% these prevent an unwanted space from occurring between the last author name
% and the end of the author line. i.e., if you had this:
% 
% \author{....lastname \thanks{...} \thanks{...} }
%                     ^------------^------------^----Do not want these spaces!
%
% a space would be appended to the last name and could cause every name on that
% line to be shifted left slightly. This is one of those "LaTeX things". For
% instance, "\textbf{A} \textbf{B}" will typeset as "A B" not "AB". To get
% "AB" then you have to do: "\textbf{A}\textbf{B}"
% \thanks is no different in this regard, so shield the last } of each \thanks
% that ends a line with a % and do not let a space in before the next \thanks.
% Spaces after \IEEEmembership other than the last one are OK (and needed) as
% you are supposed to have spaces between the names. For what it is worth,
% this is a minor point as most people would not even notice if the said evil
% space somehow managed to creep in.

% The paper headers
\markboth{IEEE TRANSACTIONS ON CONTROL SYSTEMS TECHNOLOGY, VOL -, NO. -, -}%
% \markboth{IEEE TRANSACTIONS ON POWER ELECTRONICS, VOL -, NO. -, -}%
{Shell \MakeLowercase{\textit{et al.}}: Bare Demo of IEEEtran.cls for IEEE Journals}
% The only time the second header will appear is for the odd numbered pages
% after the title page when using the twoside option.
% 
% *** Note that you probably will NOT want to include the author's ***
% *** name in the headers of peer review papers.                   ***
% You can use \ifCLASSOPTIONpeerreview for conditional compilation here if
% you desire.

% If you want to put a publisher's ID mark on the page you can do it like
% this:
%\IEEEpubid{0000--0000/00\$00.00~\copyright~2015 IEEE}
% Remember, if you use this you must call \IEEEpubidadjcol in the second
% column for its text to clear the IEEEpubid mark.

% use for special paper notices
%\IEEEspecialpapernotice{(Invited Paper)}

% make the title area
\maketitle

% As a general rule, do not put math, special symbols or citations
% in the abstract or keywords.
\begin{abstract}
In this paper we describe the design and implementation of a current controller 
for a reluctance synchronous machine  based on continuous set nonlinear model predictive control. 
A computationally efficient grey box model of the flux linkage map
is employed in a tracking formulation which is implemented 
using the high-performance framework for nonlinear model predictive control 
{\normalfont\texttt{acados}}. The resulting controller is validated in 
simulation and deployed on a {\normalfont\texttt{dSPACE}} real-time system connected to a physical reluctance synchronous 
machine. Experimental results are presented where the proposed 
implementation can reach sampling times in the range typical for electrical 
drives and can achieve large improvements in terms of control performance with respect to state-of-the-art classical control strategies.
\end{abstract}

% Note that keywords are not normally used for peerreview papers.

% For peer review papers, you can put extra information on the cover
% page as needed:
% \ifCLASSOPTIONpeerreview
% \begin{center} \bfseries EDICS Category: 3-BBND \end{center}
% \fi
%
% For peerreview papers, this IEEEtran command inserts a page break and
% creates the second title. It will be ignored for other modes.
% \IEEEpeerreviewmaketitle

\section{Introduction}

\IEEEPARstart{I}{n} recent years, reluctance synchronous machines (RSMs)
 have emerged as a competitive alternative to classical synchronous machines
 (SMs) with permanent magnet (PMSMs) or direct current excitation. In addition
 to the favourable properties of SMs in general, e.g., high efficiency, reliability
 and compact design, RSMs are often easier to manufacture and comparably cheap due to
 the absence of magnets. Moreover, their anisotropic magnetic path in the rotor, makes
 them particularly suitable for saliency-based encoderless 
 control~\cite{Inproceedings_Landsmann2010_FundamentalSaliencybasedEncoderlessControlforReluctanceSynchronousMachines,Inproceedings_Landsmann2010_ReducingtheparameterdependencyofEncoderlessPredictiveTorqueControlforreluctancemachines}. \\
However, a major drawback of the RSM concerning control is its characteristic 
nonlinearity of the flux linkage, caused by magnetic saturation and cross-coupling 
effects in the rotor. As a consequence, the machines' inductances vary significantly
with the stator currents. Moreover, additional coupling between the stator $\rm d$- and $\rm q$-currents
is imposed by the cross-coupling inductances and the coupling of the nonlinear back 
electro-motive force in the synchronous reference frame, which generally requires further
measurements to be carried out online.
\\
Regarding the control of RSMs, two main concepts have been pursued in the past:
(i) Direct Torque Control (DTC)~\cite{Article_Boldea1991_Torquevectorcontrol(tvc)ofaxially-laminatedanisotropic(ala)rotorreluctancesynchronousmotors, Article_Lagerquist1994_Sensorless-controlofthesynchronousreluctancemotor}
and (ii) field-oriented control
(FOC)~\cite{Inproceedings_Matsuo1993_Fieldorientedcontrolofsynchronousreluctancemachine, Article_Xu1991_VectorControlofaSynchronousReluctanceMotorIncludingSaturationandIronLosses,Article_Betz1993_Controlofsynchronousreluctancemachines}.
While DTC is known for its robustness and fast dynamics~\cite{Article_Bolognani2011_OnlineMTPAControlStrategyforDTCSynchronous-Reluctance-MotorDrives}, it produces a high current distortion leading to torque
ripples~\cite{Article_Chikhi2010_Acomparativestudyoffield-orientedcontrolanddirect-torquecontrolofinductionmotorsusinganadaptivefluxobserver}. In contrast, vector control improves the torque
response~\cite{Inproceedings_Rashad2004_Amaximumtorqueperamperevectorcontrolstrategyforsynchronousreluctancemotorsconsideringsaturationandironlosses} and the efficiency of the system~\cite{Article_Kamper1996_DirectFiniteElementDesignOptimisationoftheCagelessReluctanceSynchronousMachine}, but good knowledge of the system parameters is required for implementation. In~\cite{Inproceedings_Hackl2015_CurrentPI-funnelcontrolwithanti-windupforsynchronousmachines}, a completely parameter-free adaptive PI controller is proposed which guarantees tracking with prescribed transient accuracy. The controller is applied to current control of (reluctance) synchronous machines, but measurement results are not provided.
\\
In~\cite{Inproceedings_Rashad2004_Amaximumtorqueperamperevectorcontrolstrategyforsynchronousreluctancemotorsconsideringsaturationandironlosses} and~\cite{Inproceedings_Yamamoto2009_Maximumefficiencydrivesofsynchronousreluctancemotorsbyanovellossminimizationcontrollerconsideringcross-magneticsaturation}, the inductances are tracked online in order to adjust the current \emph{references} thus achieving a higher control accuracy. In~\cite{Hackl2016}, a FOC control scheme is proposed, where the PI control parameters are continuously adapted to the actual system state, which improves the overall current dynamics.  %In~\cite{Article_Morales-Caporal2007_APredictiveTorqueControlfortheSynchronousReluctanceMachineTakingIntoAccounttheMagneticCrossSaturation}, a predictive torque controller is proposed which takes into account the magnetic cross saturation of the reluctance synchronous machine. 
\par 
An alternative to classical control approaches is the use of optimization-based control techniques such as model predictive control (MPC). 
When using MPC, a parametric optimization problem is formulated that exploits a model of the plant to be controlled and enforces constraints 
while minimizing a certain objective function. Although MPC can in principle improve the control performance and ease the controller design 
\cite{Article_Geyer2009_ModelPredictiveDirectTorqueControl--PartI:ConceptAlgorithmandAnalysis}, meeting the required sampling times is 
in general a challenging task due to the high computational burden associated with the solution of the underlying optimization problems. 
\par
In order to circumvent this difficulty, several algorithmic strategies have been proposed over the past decade that use different approaches and (potentially)
different formulations of the optimal control problems to be solved. Among the possible classifications of methods present in the literature, in the fields of electrical drives and power electronics, a fundamental distinction 
can be made between what is sometimes referred to as \textit{finite} (FS-) and \textit{continuous control set} (CS-) MPC \cite{Quevedo2019}, 
\cite{Article_Cortes2008_PredictiveControlinPowerElectronicsandDrives}. 
\par
In FS-MPC, the switch positions 
of the power converter are regarded as optimization variables leading to mixed-integer programs. 
\begin{figure*}[!t]
\centering
\begin{tabular}{cc}
\subfloat[fitted grey box flux model - $\rm d$-component]{
\centering
\includegraphics[scale=0.83]{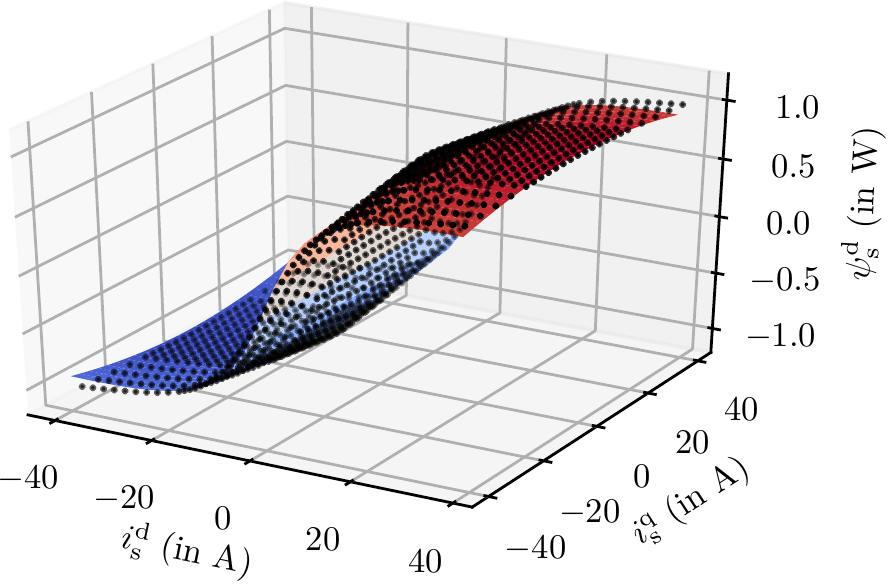}}
 \qquad
 &
 \qquad
 \qquad
\subfloat[fitted grey box flux model - $\rm q$-component]{
\centering
\includegraphics[scale=0.83]{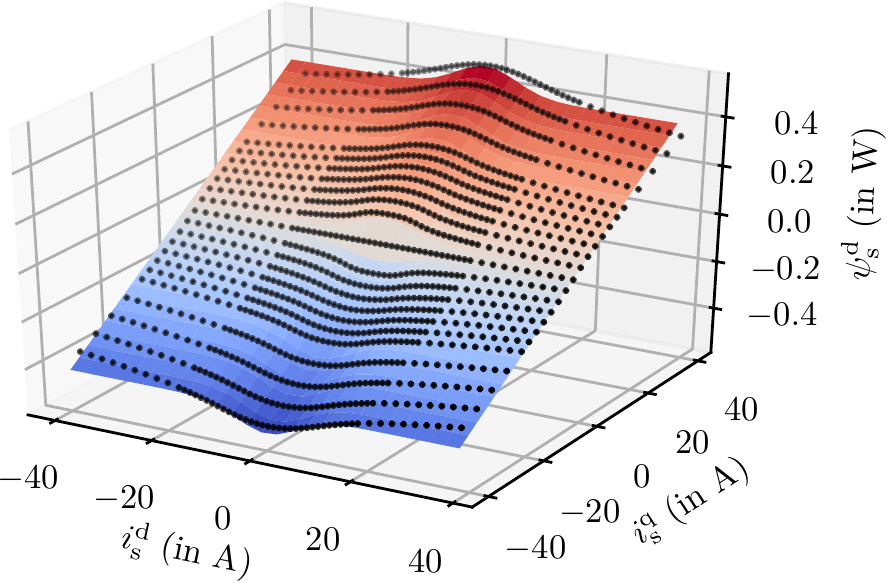}}
\end{tabular}
\caption{Nonlinear flux linkage of a real RSM obtained from FEM data $\Psisdhat$ (solid surface) and fitted grey box model $\Psi_{\rm s}^{\rm d}$ (dotted). 
The worst-case relative error amounts to less than $10 \%$.} \label{fig:fit}%\SI{9600}{\watt}
\end{figure*}
In this way, the need for an external modulator is eliminated and 
the switching sequences are directly determined by the solution to the optimal control problem (hence the name \textit{``direct''} MPC used in some of the literature on MPC 
for electrical drives and power converters \cite{Geyer2016}). 
\par
When using CS-MPC instead, we delegate the determination of switching sequences to an external modulator
in order to obtain a continuous optimization problem. For this reason, CS-MPC is sometimes referred to as \textit{``indirect''} MPC \cite{Geyer2016}. 
Although the computation times associated with this latter approach scale favourably with prediction horizon length 
and number of control variables (typically complexity $\mathcal{O}\big(N \dot (n_u + n_x)^3\big)$ can be achieved, where $N$, $n_u$ and $n_x$ represent horizon length, number 
of inputs and states, respectively), for short horizons, strategies based, e.g., on sphere decoding algorithms
applied to FS-MPC formulations can achieve sufficiently short computation times. On the contrary, CS-MPC is generally regarded as more computationally 
expensive and it is still, arguably for this reason, largely unexplored \cite{Geyer2016}. \par
Among the experimental results in the literature obtained 
with CS-MPC, in \cite{Besselmann2015} a DC-excited synchronous motor is controlled using the real-time iteration method. In \cite{Englert2018a}, a fixed-point 
iteration scheme is used to control a permanent magnet synchronous machine. 
Among applications leveraging linear-quadratic CS-MPC we mention 
the work in \cite{Domahidi2012a} in which permanent magnet synchronous machines and induction machines are controlled using explicit model predictive control. Finally, in the recent work in \cite{Cimini2020}, an active-set algorithm is 
used to solve the convex QPs arising from a linear-quadratic CS-MPC formulation to control 
a PMSM.
 
\subsection{Contribution}
In this paper, we describe the design and implementation details 
together with simulation and experimental results of a nonlinear CS-MPC controller (CS-NMPC) for an
RSM. The contributions of the present work are: 
\begin{itemize}
    \item We describe the design and implementation details of a tracking CS-NMPC 
        formulation that relies on the software package \texttt{acados}, which is capable 
        of achieving timings in the microsecond time scale necessary to 
        control the electrical drive.
    \item We propose the use of a simple grey box model for the flux maps 
        of RSMs with a small number of parameters that can be used for 
        online applications where computation times are of key importance.
    \item Finally, we present simulation and experimental results that confirm 
        the validity of the proposed control formulation and its implementation 
        and its superior performance in comparison with state-of-the-art methods
        from the field of classical control. This is, to the best of the authors' knowledge,
        one the of the earliest experimentally validated applications of CS-NMPC to 
        an RSM.
\end{itemize}

\section{Background on RSMs and NMPC}\label{sec:background}
In order to facilitate the discussion of the design and implementation of
the proposed controller, in the following, mathematical models of RSMs
and voltage source inverters (VSI) will be derived and numerical methods
for NMPC will be introduced. Note that the argument $(t)$, used to denote dependence on time,
is sometimes dropped for the sake of readability.
Moreover, we use the notation $\| x \|_P = \sqrt{x^{\top}Px}$, for some positive definite matrix $P$, to denote the $P$-weighted Euclidean norm of the vector $x$.

\subsection{Generic model of the RSM}\label{subsec:rsm}
The machine model in the synchronously rotating $(\rm d, \rm q)$-reference frame is given by~\cite[Chap.~14]{Hackl2017}
% Benutze IEEEeqnarray statt equation für Befehl \IEEEeqnarraynumspace (sonst passt die Gleichungsnummer nicht mehr neben die Gleichung
\begin{IEEEeqnarray}{l}
\begin{split}
	\usk 					 & =  \Rs \isk \!+ \! \omegak\overbrace{\begin{bmatrix}0&-1\\1&0\end{bmatrix}}^{=:\J} \psisk\big(\isk\big) \!+  \!\ddtsmall \psisk\big(\isk\big),\\
	\ddtsmall \omegak  & =  \frac{\np}{\Theta} \Big[ \mmotor(\isk) - \, \mload \Big], \qquad \ddtsmall \phik = \omegak,
\end{split}\quad
\label{eq:nonlinear RSM model}
\end{IEEEeqnarray}
where $\usk := (\usd, \, \usq)^\top$ are the applied stator voltages,
$\Rs$ is the stator resistance, $\isk := (\isd, \, \isq)^\top$ are the
stator currents and $\psisk := (\psisd, \, \psisq)^\top$ are the stator
flux linkages (functions of $\isk$). The $(\rm d, \rm q)$-reference frame rotates with
electrical angular velocity  $\omegak = \np \,\omegam$ of the rotor, where
$\np$ is the number of pole pairs and $\omegam$ denotes the mechanical
angular velocity of the machine. Furthermore, $\Theta$ is the total
moment of inertia,
\begin{equation}
  \mmotor(\isk):=\tfrac{3}{2} \np \, (\isk)^\top \J \psisk\big(\isk\big)
\end{equation}
is the electro-magnetic machine torque, and   $\mload$ represents an
external (time-varying) bounded load torque. 
\par
% \begin{figure*}[t]
%\begin{tabular}{cc}
%\subfloat[$\rm d$-component $\psisd$ of the stator  flux linkage.]{
%         \centering
%             \setlength{\figurewidth}{6cm}
%             \setlength{\figureheight}{4cm}
%	     \includetikz{\myfigurespath/tikz/}{RSM_psisd_9600W_zoom.tikz}}
%%
%\qquad
%&
%\qquad
%%
%\subfloat[$\rm q$-component $\psisq$ of the stator flux linkage.]{
%         \centering
%             \setlength{\figurewidth}{6cm}
%             \setlength{\figureheight}{4cm}
%	     \includetikz{\myfigurespath/tikz/}{RSM_psisq_9600W_zoom.tikz}}
%\end{tabular}
% \caption{Nonlinear flux linkage of a real RSM (obtained from FEM data).}%\SI{9600}{\watt}
% \end{figure*}
In order to formulate an optimal control problem, the flux dynamics can
be described, based on~\eqref{eq:nonlinear RSM model}, by the following
differential algebraic equation (DAE): 
% {\color{blue} Vector notation preferred?}
\begin{equation}\label{eq:DAE}
    \begin{aligned}
        &&\ddtsmall\psisk &= \usk-\Rs\isk-\omegak\J\psisk + \vk, \\
        &&0               &= \psisk - \Psisk(\isk), 
    \end{aligned}
\end{equation}
%
% {\color{red} Component-wise notation preferred?}
% \begin{equation}
%     \begin{aligned}
%         &&\ddtsmall\psisd &= \usd-\Rs\isd+\omegak\psisq + \vd, \\
%         &&\ddtsmall\psisq &= \usq-\Rs\isq-\omegak\psisd + \vq, \\
%         &&0            &= \psisd - \Psisd(\isd, \isq), \\
%         &&0            &= \psisq - \Psisq(\isd, \isq),
%         % &&0            &= \tilde{\psi}_{\rm d}(i_{\rm d}, i_{\rm q}), \\
%         % &&0            &= \tilde{\psi}_{\rm q}(i_{\rm d}, i_{\rm q}),
%     \end{aligned}
% \end{equation}
where $\Psisk := (\Psisd, \, \Psisq)^\top \vcentcolon \R^2 \rightarrow \R^2$ 
% {\color{red}$\Psisd \vcentcolon \R^2 \rightarrow \R$ and $\Psisq \vcentcolon \R^2 \rightarrow \R$} 
defines the algebraic constraints based on the identified flux maps and 
$\vk := (\vd, \, \vq)^\top$ 
% {\color{red}$\vd, \vq$}
are additive disturbances which will be used in an offset-free NMPC setting (see Section \ref{subsec:formulation}). 

Based on the available flux maps computed through the finite element method (FEM), 
we obtained a continuously differentiable model by fitting
a simple grey box model. Due to their low number of parameters and simple structure, 
we propose the following parametrization of the flux maps:
\begin{equation}
    \begin{aligned}
        &&\Psisd(\isd, &\isq, \ThetaVecd) =  \\
        && &\tfrac{c^{\rm d}_0}{\sqrt{2\pi\sigma_{\rm q}^2}} \exp \left(-\gamma\left(\isq, \sigma_{\rm q}\right)\right)
        \mathrm{atan}(c^{\rm d}_1 \, \isd) + c^{\rm d}_2 \, \isd\\
    \end{aligned}
\end{equation}
and
\begin{equation}
    \begin{aligned}
        &&\Psisq(\isd&, \isq, \ThetaVecq) =  \\
        && &\tfrac{c^{\rm q}_0}{\sqrt{2\pi\sigma_{\rm d}^2}} \exp \left(-\gamma\left(\isd, \sigma_{\rm d}\right)\right)
        \mathrm{atan}(c^{\rm q}_1 \, \isq) + c^{\rm q}_2 \, \isq,\\
    \end{aligned}
\end{equation}
with 
\begin{equation}
    \gamma(x,y) \vcentcolon = \frac{1}{2}\left(\frac{x}{y}\right)^2
\end{equation}
and where the unknown parameters involved are 
\begin{equation}\ThetaVecd \vcentcolon = (c^{\rm d}_0,\, c^{\rm d}_1,\, c^{\rm d}_2,\,\sigma_{\rm d})
\end{equation} and
\begin{equation}\ThetaVecq \vcentcolon = (c^{\rm q}_0,\, c^{\rm q}_1,\, c^{\rm q}_2,\,\sigma_{\rm q}).\end{equation}
% The main intuition behind such a model lies in the fact that the flux map for the $\rm d$ component, resemble 
This parametrization of the flux maps is, to the authors' best knowledge, novel and 
it is able to capture the main features of the flux maps with only 4 parameters 
per flux component.
The numerical values of the coefficients can be computed by solving the 
following (decoupled) nonlinear least-squares
problems: 
\begin{equation}
	\begin{split}
        \underset{\ThetaVecd}{\min}	\,\,  &\sum_{j=1}^{m}\sum_{k=1}^{n} \left(\Psisd(\isdbar{j}, \isqbar{k}, \ThetaVecd) - \Psisdhat(\isdbar{j}, \isqbar{k})\right)^2\\ 
    \underset{\ThetaVecq}{\min}	\,\,  &\sum_{j=1}^{m}\sum_{k=1}^{n} \left(\Psisq(\isdbar{j}, \isqbar{k}, \ThetaVecq) - \Psisqhat(\isdbar{j}, \isqbar{k})\right)^2\\ 
	\end{split}
\end{equation}
where $\isdbar{j}$ and $\isqbar{k}$ are the $j$-th and $k$-th current 
data points associated with the flux values 
$\Psisdhat$ and $\Psisqhat$ obtained from FEM analysis.
The fitting problems have been solved with the \texttt{MATLAB} Curve Fitting Toolbox and the 
resulting fitted model is shown in Figure~\ref{fig:fit}.%~and~ \ref{fig:fit_error}, respectively. 
\begin{figure}[t]
	\centering
    \hspace{0.5cm}
	\vspace{-0.6cm}
    \par
    \includegraphics[scale=1.0]{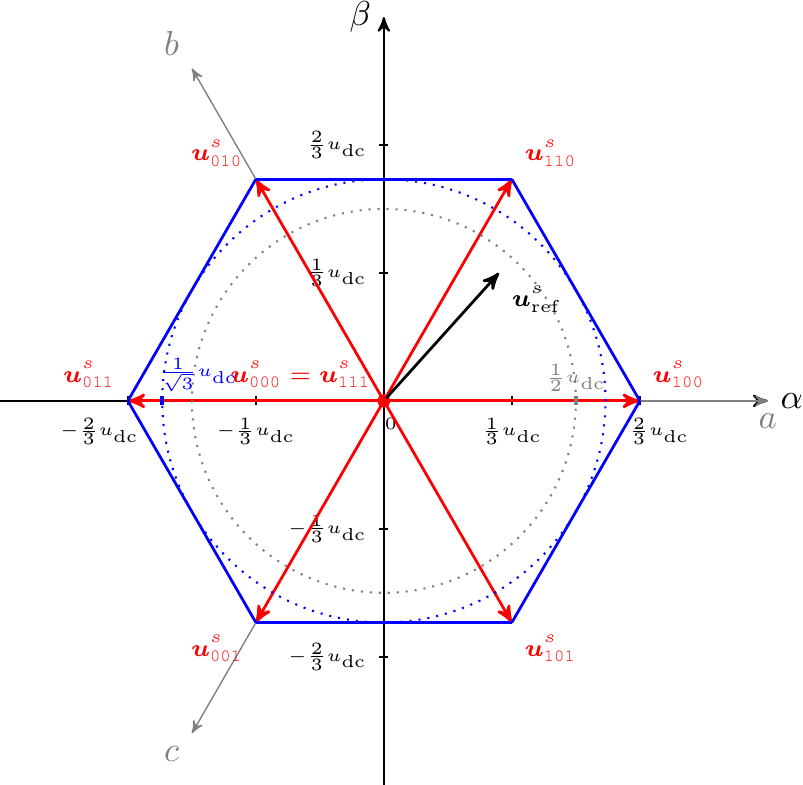}
	\caption{Voltage hexagon associated with the two-level VSI.}
	\label{fig:voltage_hexagon_2l}
\end{figure}

\subsection{Model of the two-level VSI}\label{subsec:VSI}
The machine is supplied by a two-level voltage source inverter (VSI), which -- 
on average over one switching period $\Ts$ -- translates a given voltage 
reference 
\begin{equation}
\ussref \vcentcolon = (u_{s, \textrm{ref}}^{\alpha}, u_{s, \textrm{ref}}^{\beta})
\end{equation}
(in the stationary $s = (\alpha, \beta)$-reference frame) 
into the inverter output voltage $\uss$, i.e.
\begin{equation}
	\uss(k\,\Ts) \approx \ussref((k-1)\Ts), \quad k \in \N.
\end{equation}
Since a two-level voltage source inverter may produce a total of eight 
unique switching vectors, i.e.  $\ssabc := (\ssa, \ssb, \ssc)^\top \in\{000, 001, 010, 100, 011, 101, 110, 111\}$, the typical voltage hexagon in the $\alpha\beta$-plane is obtained (see Figure~\ref{fig:voltage_hexagon_2l}), 
where
\begin{equation}
\uss = \kappa\, \udc  \begin{bmatrix} \tfrac{1}{2} & 0 & -\tfrac{1}{2} \\ 0 & \tfrac{\sqrt{3}}{2} &
0 \end{bmatrix}  \begin{bmatrix} 1 & -1 & 0 \\ 0 & 1 & -1 \\ -1 & 0 & 1\end{bmatrix} 
\ssabc
\end{equation}
depends on the switching vector $\ssabc$ and the Clarke-factor $\kappa \in \{2/3, \sqrt{2/3}\}$
\cite[Chap. 14]{Hackl2017}. %The switching vector is obtained by pulse width modulation (PWM) or space vector modulation (SVM).
Using space-vector modulation (SVM) to generate the switching vector, 
any voltage reference within the circle of radius $\udc/\sqrt{3}$ can be 
realized, with $\udc$ denoting the (assumed constant) DC-link voltage. 
Finally, the inverter output voltage is transformed into the rotating 
$(\rm d, \rm q)$-reference frame using the inverse Park transformation, i.e.

\begin{equation}
	\usk = \begin{bmatrix} \usd \\ \usq \,\,\end{bmatrix} = 	
	\underbrace{\begin{bmatrix}
		\cos(\phik) & \sin(\phik)\\
		-\sin(\phik) & \cos(\phik)
	\end{bmatrix}}_{=: \Tp(\phik)^{-1}} \uss.
\end{equation}

From now on, since we will only refer to currents, fluxes and voltages applied to the stator and 
in the ($\rm d$-$\rm q$)-frame, we will simplify the notation by dropping the associated subscript such 
that, for example, $i = (\isd, \isq)$ denotes the stator currents in the ($\rm d$-$\rm q$)-frame.
%\label{fig:flux data of real RSM (FEM)}
%\begin{figure*}[t]
%\begin{tabular}{cc}
%\subfloat[fitting error $\Psisdhat - \Psisd$]{
%\centering
%\includegraphics[scale=0.8]{Figures/pdf/fit_data_error_d.pdf}}
%\qquad
%&
%\qquad
%%
%\subfloat[fitting error $\Psisqhat - \Psisq$]{
%\centering
%\includegraphics[scale=0.8]{Figures/pdf/fit_data_error_q.pdf}}
%\end{tabular}
%\caption{Flux maps fitting error for grey box model.} \label{fig:fit_error}%\SI{9600}{\watt}
%\end{figure*}

\subsection{Nonlinear model predictive control}
NMPC is an optimization-based control strategy that allows one to tackle
control problems involving potentially nonlinear dynamics, constraints and objectives 
by solving online a series of parametric nonlinear programs (NLP). Due 
to the computational challenge of solving NLPs within the required sampling times, 
NMPC has initially found application in the chemical industry and in the 
field of process control in general \cite{Rawlings2017}, where relatively slow dynamics allow 
for sufficiently long sampling times. In more recent times, due to 
the development of increasingly efficient numerical methods and 
software implementation and due to the growing computational 
power of embedded control units, NMPC has gradually become 
a viable approach for applications with much shorter computation times. Among other 
recent works that reported on the successful application of MPC to control systems 
with sampling times in the range of milli- and microsecond we mention \cite{Zanelli2018, Albin2017}.
\par 

\begin{figure}[t]
	\centering
  % \begin{minipage}{5.25cm}
  % \subfloat[Real-time system (A) and inverters (B1,B2).]{%
	% \includegraphics{Figures/pdf/CRES_dSPACE_Schaltschrank_small.pdf}
  %   }
  % \end{minipage}
  %  \hfill
  % \begin{minipage}{10cm}
  %   \subfloat[Host-PC (C) for rapid-prototyping and data acquisition.]{%
  % \includegraphics{Figures/pdf/CRES_HostPC_Sicherheitsbox_small.pdf}
  %   }
  %     \vfill
    % \subfloat[Reluctance (D1) and permanent-magnet (D2) synchronous machine with torque sensor.]{%
    % \includegraphics{Figures/pdf/CRES_RSM_PMSG_small.pdf}
		\includegraphics[scale=0.225]{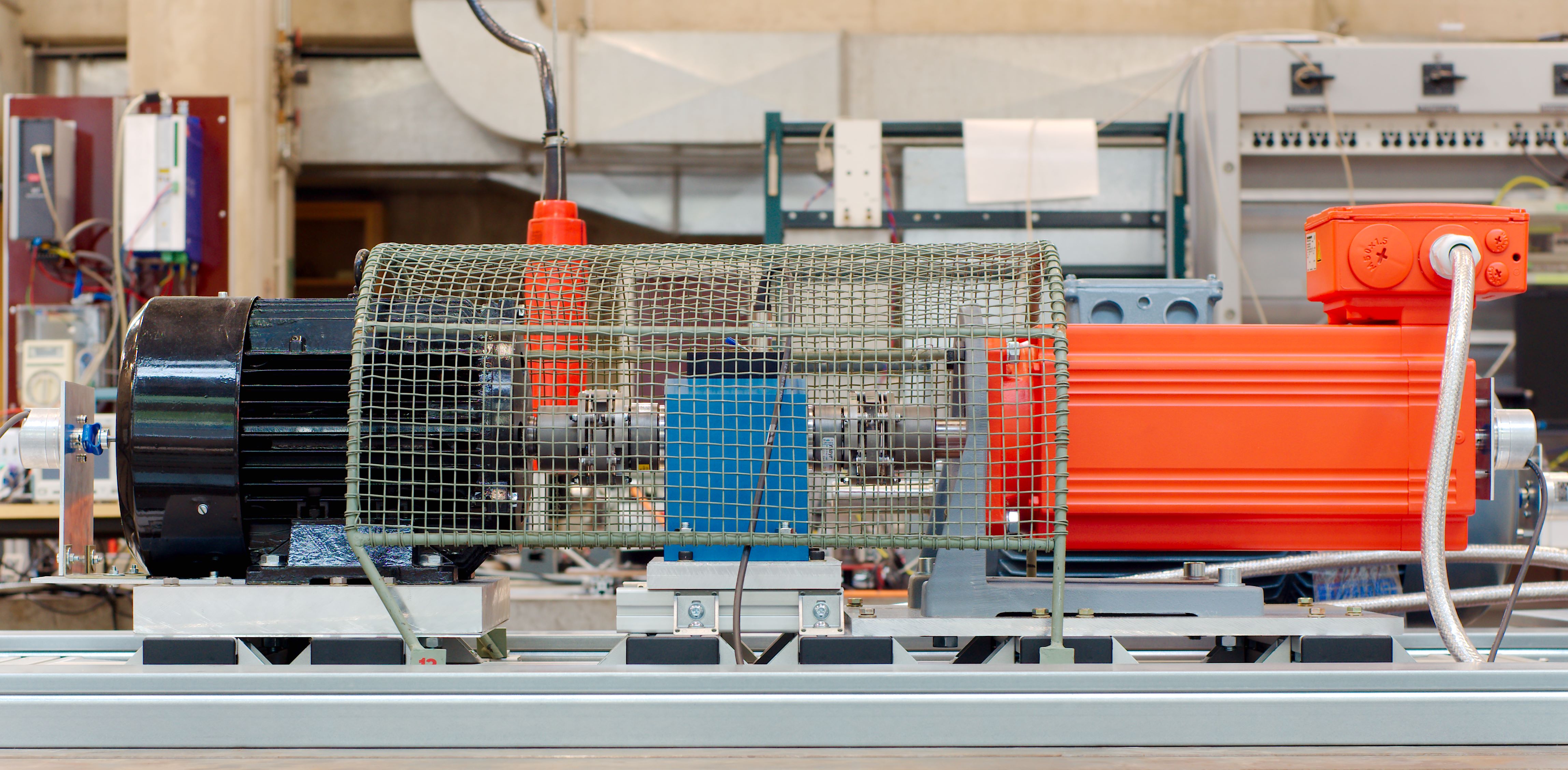}
    % }
  % \end{minipage}
        \caption{Laboratory setup with \texttt{dSPACE} real-time system, voltage-source inverters connected back-to-back, RSM, PMSM and torque sensor.}
 \label{fig:lab_setup}
\end{figure}
\begin{figure}[t]
	\centering
	\includegraphics{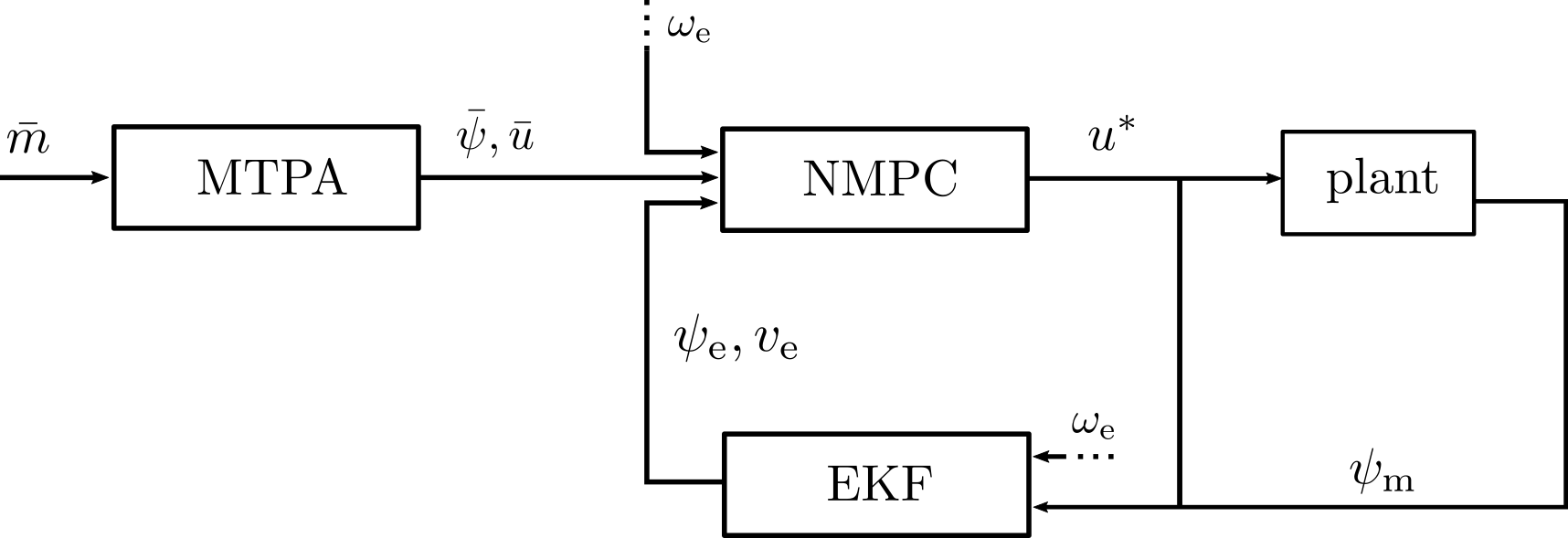}
    \caption{Control diagram: the MTPA LUTs provide the reference flux $\bar{\psi}$ and voltage $\bar{u}$ 
    associated with a given reference torque $\bar{m}$. The NMPC controller computes the optimal control action based on 
    the current state and disturbance estimate provided by an EKF. }
    \label{fig:control_diagram}
\end{figure}

In this paper, we will regard the following standard tracking formulation with prediction horizon 
$T_h$ and $N$ shooting nodes,
where the squared deviation of fluxes $\psi$ and voltages $u$ from properly 
defined steady-state references are penalized:
\begin{equation}\label{eq:tracking_nmpc}
{\!\!\!\!\!}{\!\!}\begin{aligned}
&\underset{\begin{subarray}{c}
\psi_0, \dots, \psi_N \\
u_0, \dots, u_{N-1}
\end{subarray}}{\min}	    &&\!\!\!\frac{T_h}{2N}\sum_{i=0}^{N-1} \bigg\| \begin{matrix}\psi_i - \bar{\psi} \\ u_i - \bar{u}\end{matrix} \,\bigg\|_W^2 \!\!\!+ \frac{1}{2}\| \psi_N - \bar{\psi} \|^2_{W_N} \\ 
&\,\,\,\quad \text{s.t.}    &&\!\!\!\psi_0 - \psi_{\mathrm{e}} = 0, \\
& 						    &&\!\!\!g(\psi_i,u_i, \omega_{\text{e}}, v_{\text{e}}) - \psi_{i+1}\! = 0,  \,\,i = 0,\dots, N-1,\\
& 						    &&\!\!\!u_i^{\top}u_i \leq \left(\frac{u_{\text{dc}}}{\sqrt{3}}\right)^2, \quad  \quad \quad \,\,\,\,\,\,\,\,\, i = 0,\dots, N-1,\\
& 						    &&\!\!\!\hat{C}u_i \leq \hat{c}, \quad \quad \quad \quad \quad  \quad \quad \,\,\,\,\,\,\, i = 0,\dots, N-1,\\
\end{aligned}{\!\!\!}
\end{equation}
where $g$ describes the discretized dynamics obtained by integrating the differential-algebraic model 
in \eqref{eq:DAE} using the Gauss-Legendre collocation method of order 2 
assuming constant (estimated) angular velocity $\omega_{\text{e}}$ and disturbances $v_{\text{e}}$. The variables 
$\bar{\psi}$ and $\bar{u}$ denote the steady-state references computed 
for a given desired torque using a maximum-torque-per-Ampere (MTPA) criterion \cite{Eldeeb2019}.
Given the flux maps obtained from FEM data in Figure \ref{fig:fit}, it is possible 
to compute off-line lookup tables (LUTs) that contain the MTPA reference fluxes and voltages for 
a finite number of values of the target torque in a specified range. The LUTs are then interpolated 
online in order to compute approximate values of $\bar{\psi}$ and $\bar{u}$ associated with the 
specified target torque $\bar{m}$ (see Figure \ref{fig:control_diagram}).

\begin{figure*}[t]
\centering
\begin{tabular}{cc}
\hspace{-0.6cm}
\subfloat{
    \includegraphics[scale=0.81]{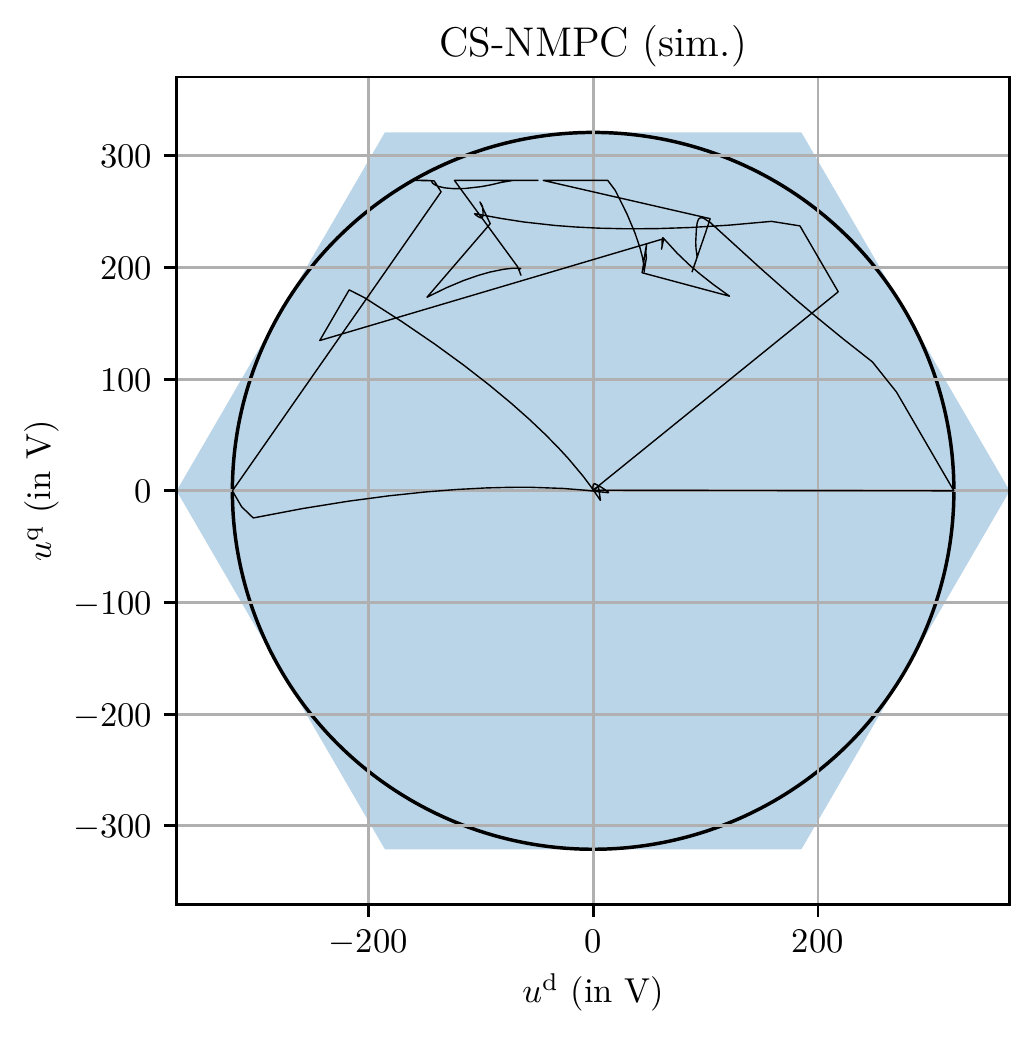}}
	% \qquad 
	& 
	% \qquad
\subfloat{
\includegraphics[scale=0.81]{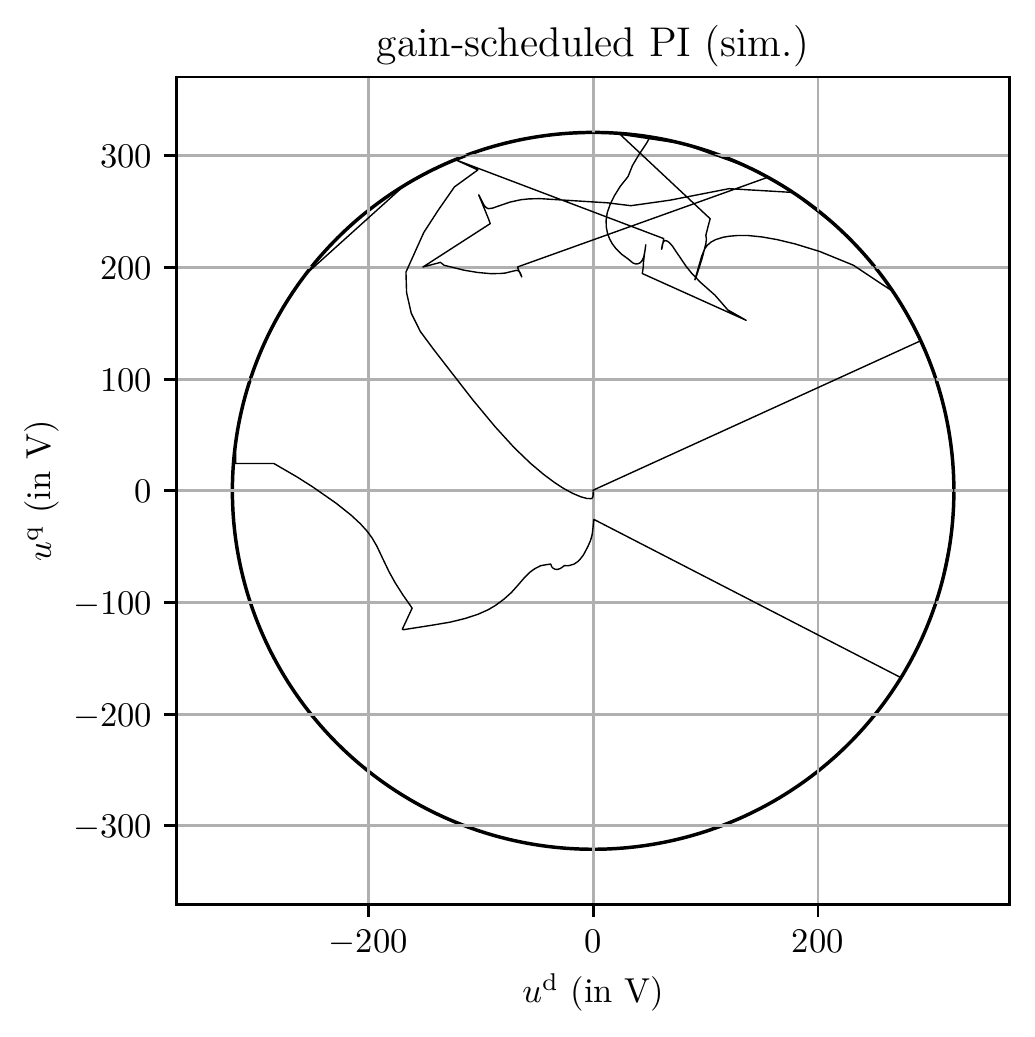}}
\end{tabular}
\caption{Current steps at $\SI{157}{\radian\per\second}$ (simulation): results obtained
	using the CS-NMPC (left) and the gain-scheduled PI controller (right). The voltage 
    spherical constraints are directly included into the control formulation
	using the SCQP strategy proposed in \cite{Verschueren2016}. Additionally, a ``safety'' 
	polytopic constraint is included which, due to its linearity, is always 
	satisfied exactly.}\label{fig:sim_u}
\end{figure*}

The convex quadratic constraint in \eqref{eq:tracking_nmpc} describes 
the circular input feasible set introduced in Section~\ref{subsec:VSI}. Finally, $\hat{C}$ and $\hat{c}$ define polytopic constraints (which we will later refer to as ``safety'' constraints) that are meant to be always inactive at any local solution 
of \eqref{eq:tracking_nmpc} (apart from a finite number of points where they are locally equivalent to the linearized
spherical constraint), but can mitigate constraint violation of intermediate SQP iterates.
In particular, we define $\hat{C}$ and $\hat{c}$ such that the affine constraint defines an outer polytopic approximation 
with 6 facets as depicted in Figure \ref{fig:sim_u}. Notice that due to this formulation of the feasible set,
linear independence constraint qualification can fail at a finite number of points where the linearization 
of the nonlinear constraint is equivalent to one of the affine constraints in \eqref{eq:tracking_nmpc}. Although, 
this would violate a common assumption used in convergence theory for both SQP and some numerical methods 
for the solution of convex QPs, the active-set solver \texttt{qpOASES} that we employ for this application can handle 
redundant constraints through a strategy that determines which constraint needs to be removed from the working set \cite{Ferreau2008}.
\begin{remark}
    Notice that the actual dynamics of the system involve a coupling of mechanical ($\omega$) and electrical 
    states ($\psi$). It is however common, given the large difference between associated time constants, 
    to assume a constant angular velocity $\omega$ when designing controllers. In our case, it allows us to 
    use much shorter prediction horizons since we do not require the OCP in \eqref{eq:tracking_nmpc} to 
    steer the speed of the motor to the desired reference, but only fluxes which directly map to currents and, 
    for a given speed, to torques.
\end{remark}
% \begin{equation}\label{eq:nmpc}
% {\!\!\!\!\!}{\!\!}\begin{aligned}
% &\underset{\begin{subarray}{c}
% s_0, \dots, s_N \\
% u_0, \dots, u_{N-1}
% \end{subarray}}{\min}	    &&\sum_{i=0}^{N-1} l(s_i, u_i) + m(s_N)\\ 
% &\,\,\,\quad \text{s.t.}    &&s_0 - x = 0, \\
% & 						    &&f(s_i,u_i)  - s_{i+1} = 0, \,\,\,\, i = 0,\dots, N-1,\\
% & 						    &&\pi(s_i, u_i) \leq 0, \quad \quad \quad \,\,\,\, i = 0,\dots, N-1,\\
% & 						    &&\pi_N(s_N) \leq 0, 
% \end{aligned}{\!\!\!}
% \end{equation}

% where $s_i \in \mathbb{R}^{n_s}$ and $u_i \in \mathbb{R}^{n_u}$ describe the predicted states and inputs of
% the system to be controlled, respectively. The functions 
% $l \vcentcolon \mathbb{R}^{n_s} \times \mathbb{R}^{n_u} \rightarrow \mathbb{R}$,  
% $m \vcentcolon \mathbb{R}^{n_s} \rightarrow \mathbb{R}$ describe the stage and 
% terminal cost, respectively and $f \vcentcolon \mathbb{R}^{n_s} \times \mathbb{R}^{n_u} \rightarrow \mathbb{R}^{n_s}$, 
% $\pi \vcentcolon \mathbb{R}^{n_s} \times \mathbb{R}^{n_u} \rightarrow \mathbb{R}^{n_{\pi}}$ and
% $\pi_N \vcentcolon \mathbb{R}^{n_s} \rightarrow \mathbb{R}^{n_{\pi_{\scaleto{N}{2pt}}}}$ describe the dynamics of the system, 
% the stage and terminal constraints, respectively. We will make the assumption that $l$, $m$, $\psi$ $\pi$ and $\pi_N$ are 
% twice continuously differentiable. Finally, the parameter $x$ denotes the current state 
% of the system. 
\par
Problem \eqref{eq:tracking_nmpc} is used to define an implicit feedback policy that requires the solution 
of an instance of the parametric NLP at every sampling time, where the value of the parameter $\psi_{\rm e}$
is given by the current estimate of the system's state. The resulting solutions are feasible with respect 
to the constraints and minimize (at least locally) the cost function. Nominal and inherently robust stability
of the closed-loop system can be guaranteed in a neighborhood of a steady-state by properly choosing the terminal 
cost \cite{Rawlings2017}.
\par
\begin{remark}
	Notice that formulations more general than \eqref{eq:tracking_nmpc} can in principle be used
	in the framework of NMPC. Among others, economic costs and more general nonlinear constraints 
	and nonlinear cost terms, are features that can be included in the problem in order to 
	better capture control design requirements. However, for the application discussed in this paper,
	the nonlinear least-squares problem described in \eqref{eq:tracking_nmpc} is general enough.
\end{remark}
% Although recent progress has been made in the field of economic MPC, we will refer in this work to a standard 
% ``tracking'' formulation, where $l$ and $m$ are assumed to be positive definite, which simplifies both 
% stability analysis and numerical methods. Moreover, a formulation with least-squares terms of the 
% form 
% \begin{equation}
%     l = \frac{1}{2}\| V_x s + V_u u  - y_{\text{ref}}\|_W^2, \quad m = \frac{1}{2}\| s - x_{\text{ref}}\|_{W_N}^2,
% \end{equation}
% can naturally encode the requirement of tracking the reference steady-state state and input.
\subsection{Numerical methods and software for NMPC}
In order to be able to solve problem \eqref{eq:tracking_nmpc} within 
the available computation time, the use of 
efficient numerical methods is fundamental. First, 
since \eqref{eq:tracking_nmpc} is obtained through a multiple-shooting 
discretization strategy, an efficient way of computing evaluations of the discretized dynamics $g$ and 
of its first-order (and potentially second-order) derivatives needs to be available. This is commonly 
achieved by means of numerical integration of the ordinary differential equation (ODE) 
or algebraic differential equations (DAE) describing the dynamics of the system. Although 
for ODEs explicit and implicit integration methods can be used, for DAEs, as for 
 the system under consideration (see Section \ref{subsec:rsm}), implicit schemes 
are generally necessary that involve the solution of nonlinear root-finding problems via Newton-type
iterations.
In the context of NMPC, efficient strategies and tailored implementations are available that rely on, for example,
exploiting the structure of the model \cite{Quirynen2013a, Frey2019}, on reuse of Jacobian factorizations and 
efficient sensitivity generation \cite{Quirynen2012a}, on lifting-based formulations \cite{Quirynen2015a} and 
on inexact iterations \cite{Quirynen2018}.
\par
Once the linearization is carried out, one generally needs to solve structured linear systems that can 
be used to compute the solution to a quadratic program (QP) as in sequential quadratic programming (SQP), compute the update 
defined by an interior-point method or the one used by other various strategies such as, e.g., first-order methods. 
Since the description of the details of the different available approaches to solve \eqref{eq:tracking_nmpc} goes well 
beyond the scope of this work, we will focus, in the following, on the SQP strategy, which constitutes the basis 
for the real-time iteration (RTI) method used in this application.
\par
When using SQP, a sequence of structured QPs of the following form needs to be solved:
\begin{equation}\label{eq:qp}
{\!\!\!}{\!\!}\begin{aligned}
&\underset{\begin{subarray}{c}
s_0, \dots, s_N \\
u_0, \dots, u_{N-1}
\end{subarray}}{\min}	    &&\frac{1}{2}\sum_{i=0}^{N-1} \begin{bmatrix} s_i \\ u_i \\ 1\end{bmatrix}^{\top}\!\!\!H_i\begin{bmatrix} s_i \\ u_i \\ 1\end{bmatrix} + \frac{1}{2}\begin{bmatrix} s_N \\ 1\end{bmatrix}^{\top}\!\!H_N\begin{bmatrix} s_N \\ 1\end{bmatrix}\\ 
&\,\,\,\quad \text{s.t.}    &&s_0 - x = 0, \\
& 						    &&s_{i+1} = A_is_i + B_iu_i + c_i, \,\, i = 0,\dots, N-1,\\
& 						    &&C_i u_i + D_i s_i + e_i \leq 0, \,\,\,\, \quad  i = 0,\dots, N-1,\\
& 						    &&D_N s_N + e_N \leq 0. 
\end{aligned}
\end{equation}
Here the parameter $x$ denotes the current state of the system and the states and inputs are denoted 
by $s \in \mathbb{R}^{n_s}$ and $u \in \mathbb{R}^{n_u}$, respectively.
The matrices and vectors $A_i \in \mathbb{R}^{n_s \times n_s}$, $B_i \in \mathbb{R}^{n_s \times n_u}$ and 
$c_i \in \mathbb{R}^{n_s}$, for $i=0,\dots,N-1$, define the linearized dynamics obtained through numerical integration and 
where $C_i \in \mathbb{R}^{n_{\pi} \times n_{u}}$, $D_i \in \mathbb{R}^{n_{\pi} \times n_{s}}$, 
$e_i \in \mathbb{R}^{n_{\pi}}$, for $i=0,\dots,N-1$, and 
$D_N \in \mathbb{R}^{n_{\pi_{\scaleto{N}{2pt}}} \times n_{s}}$, $e_N \in \mathbb{R}^{n_{\pi_{\scaleto{N}{2pt}}}}$ define 
the linearized constraints. Finally the matrices
\begin{equation}
	H_i = \begin{bmatrix} Q_i & S_i & q_i\\ S_i^{\top}& R_i & r_i \\ q_i^{\top} & r_i^{\top} & \,0\, \end{bmatrix} \quad \text{and} \quad H_N = \begin{bmatrix} Q_N & q_N \\ q_N^{\top} & 0 \end{bmatrix}, 
\end{equation}
with $Q_i\in \mathbb{R}^{n_s \times n_s}$, $S_i\in \mathbb{R}^{n_s \times n_u}$, $R_i\in \mathbb{R}^{n_u \times n_u}$, $r_i\in \mathbb{R}^{n_u}$, $q_i\in \mathbb{R}^{n_s}$, for $i=0,\dots,N-1$, and $Q_N$, $q_N$ define the cost of the QP.
The matrices and vectors defining the QP \eqref{eq:qp} are computed 
based on the linearization associated with the current primal-dual iterate 
$z = (s, u, \lambda\!\!,\, \mu)$ (where $\lambda$ and $\mu$ represent the Lagrange multipliers 
associated with the equality and inequality constraints in \eqref{eq:tracking_nmpc}, respectively) and, after solving 
\eqref{eq:qp}, the iterate is updated, i.e., $z_{+} \leftarrow z + \alpha \Delta z$,
where $\Delta z$ represents the primal-dual step associated with the solution of the QP constructed at the 
linearization point $z$ and $\alpha > 0$ is the step size, which can 
be adjusted to achieve convergence. Under standard assumptions \cite{Nocedal2006}, the iterates
 converge to a local minimum of \eqref{eq:tracking_nmpc}.
\begin{figure*}[t]    
\begin{tabular}{cc}
\hspace{-0.6cm}
\subfloat{
    \includegraphics[scale=0.82]{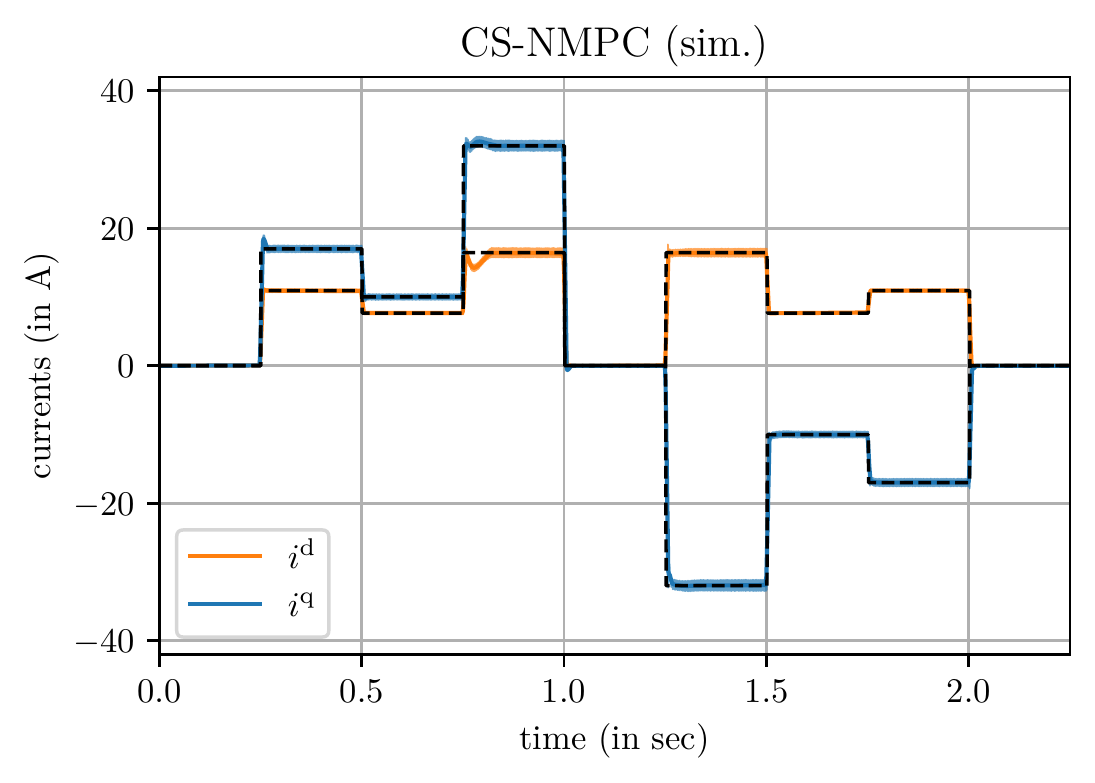}
}
&\hspace{-0.6cm}
\subfloat{
    \includegraphics[scale=0.82]{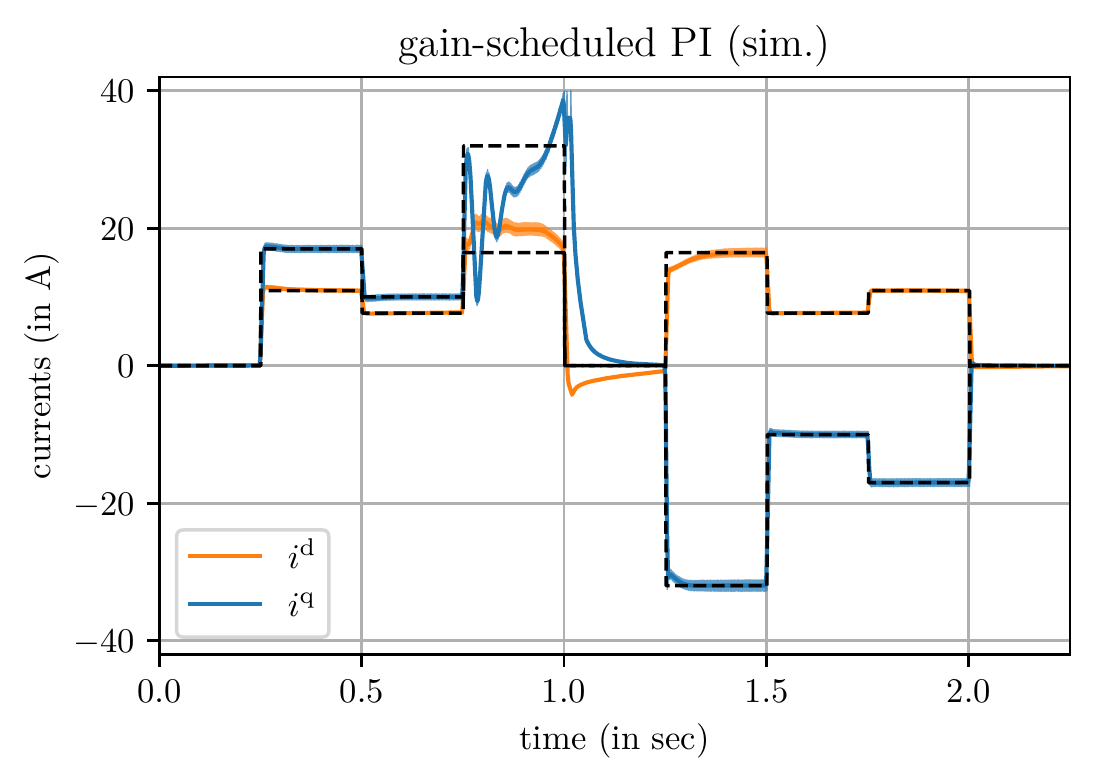}
}
\end{tabular}
\caption{Current steps at $\SI{157}{\radian\per\second}$ (simulation): results obtained using the CS-NMPC (left) and gain-scheduled PI controller (right). The CS-NMPC controller outperforms the PI controller, especially when the input constraints become active (e.g., between $t=\SI{0.75}{\second}$ and $t=\SI{1.00}{\second}$). At the same time, a faster transient can be achieved even when the constraints become active only for a short time.}\label{fig:sim_TS_157_currents}
\end{figure*}
\begin{figure*}[t]    
\begin{tabular}{cc}
\hspace{-0.6cm}
\subfloat{
\includegraphics[scale=0.82]{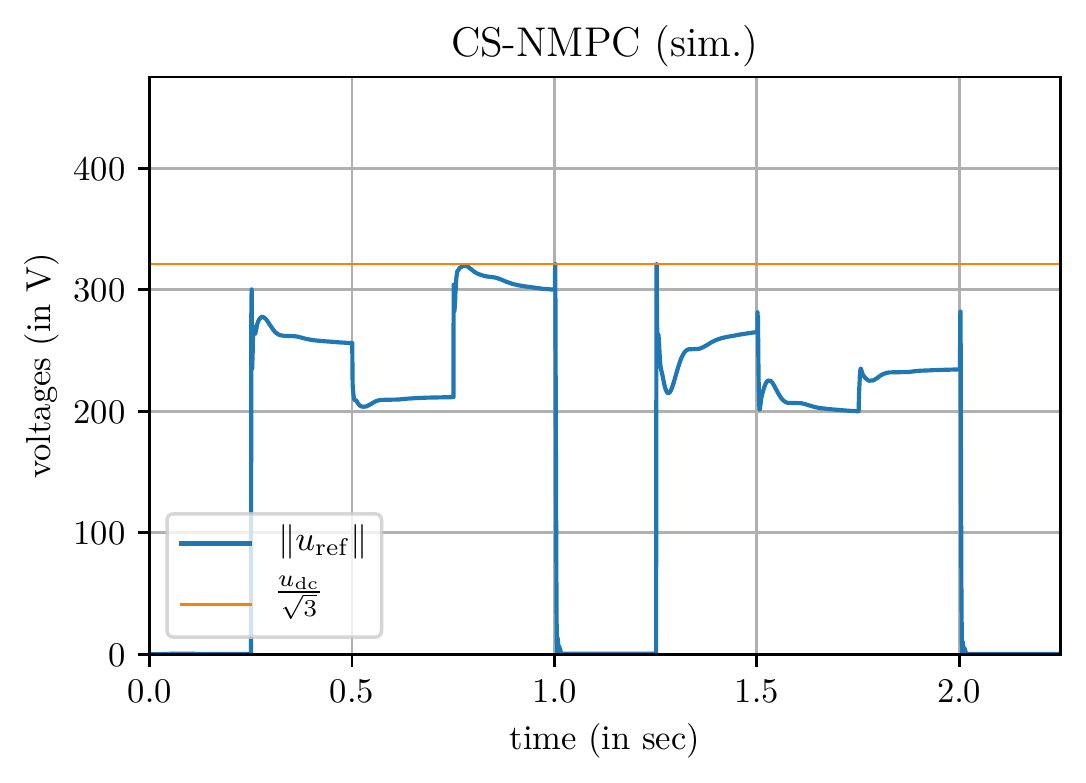}
}
&\hspace{-0.6cm}
\subfloat{
\includegraphics[scale=0.82]{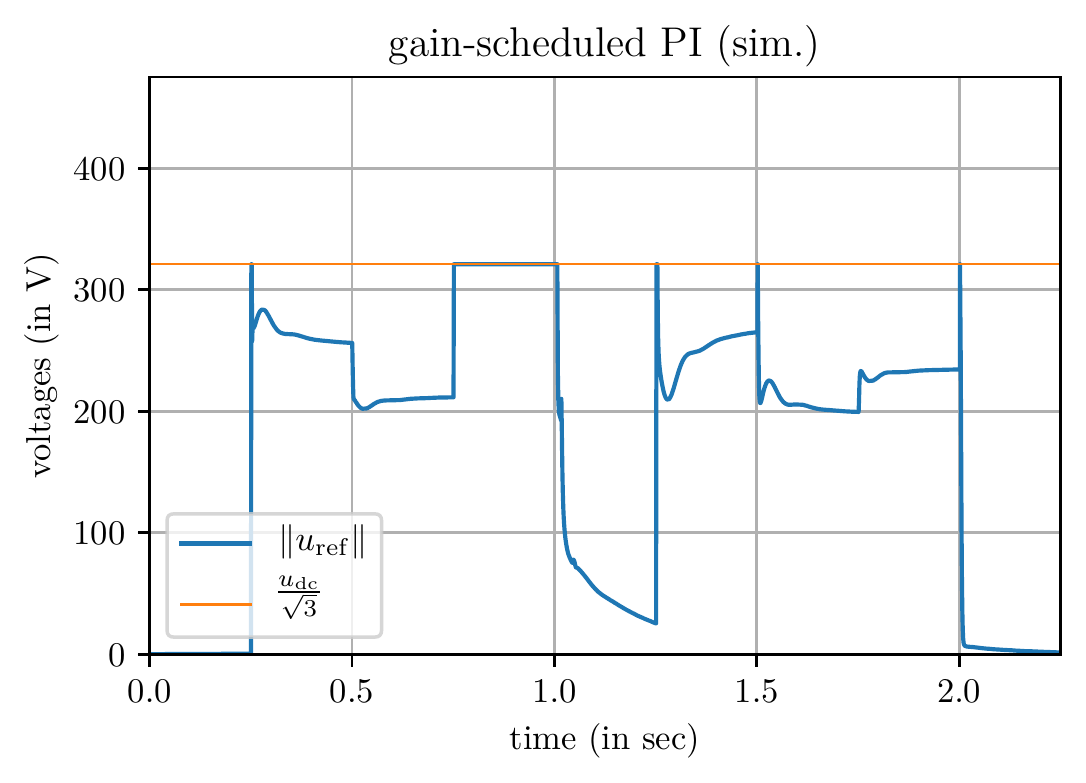}
}\end{tabular}
\caption{Current steps at $\SI{157}{\radian\per\second}$ (simulation): two-norm of voltage references $u_{\text{ref}}$ commanded by the two controllers and $u_{\text{dc}}$ over time in simulation at $\SI{157}{\rad\per\second}$ . During the third current step, the PI controller saturates and does not steer the system to the desired reference. Notice that the input commanded by the PI controller remains saturated during the entire step. On the contrary, the CS-NMPC controller, after an initial saturation, steers the current to the (feasible) reference values. }\label{fig:sim_TS_157_voltages}
\end{figure*}
% \subsection{Inexact methods for NMPC and the RTI strategy}
\par
Due to the computational burden associated with the solution of the QPs and re-linearization 
of the original NLP in \eqref{eq:tracking_nmpc}, several approximate strategies can be used that can 
significantly reduce computation times (e.g., \cite{Zanelli2019}, \cite{Graichen2010},  \cite{Feller2017}). 
In this work, we will use the RTI strategy \cite{Diehl2001, Diehl2002a}, which relies on a single SQP iteration in order
to provide an approximate feedback law. Attractivity properties of the combined system-optimizer dynamics 
associated with the RTI strategy have been first analyzed in \cite{Diehl2005b} and \cite{Diehl2007b} for a simplified setting 
without inequality constraints and the recent work in \cite{Zanelli2020} and \cite{Zanelli2020a} extends these results 
proving asymptotic stability for the inequality constrained case too. 
The RTI has been implemented in the software packages \texttt{MUSCOD-II} \cite{Diehl2001b}, 
\texttt{ACADO} \cite{Houska2011e} and in its successor \texttt{acados} \cite{Verschueren2018}, 
which is shortly described in the next subsection.

\subsection{The \texttt{acados} framework}

The high-performance software package \texttt{acados} \cite{Verschueren2018}
provides a modular framework for NMPC and moving horizon estimation (MHE). It 
consists of a C library that implements building blocks 
needed to solve NLPs arising from NMPC and MHE formulations.
It relies on the high-performance linear algebra package 
\texttt{BLASFEO} \cite{Frison2018} and on the quadratic program (QP) solver 
\texttt{HPIPM} \cite{Frison2020} and contains efficient implementations of explicit 
and implicit integration methods. Moreover, it interfaces 
a number of QP solvers such as \texttt{qpOASES} \cite{Ferreau2014}
% , \texttt{qpDUNES} \cite{Frasch2015} 
and \texttt{OSQP} \cite{Stellato2020} and it provides high-level 
Python and \texttt{MATLAB} interfaces. Through these interfaces, one can conveniently specify 
optimal control problems and code-generate a self-contained C library 
that implements the desired solver and can be easily deployed onto 
embedded control units such as \texttt{dSPACE} using the automatically 
generated C wrapper and S-Function. The code-generation takes place 
through templated C code which is rendered by the \texttt{Tera} templating 
engine written in Rust. In this way, human-readable C code can be 
generated that facilitates the deployment on the target hardware.
\subsection{NMPC offset-free tracking formulation}\label{subsec:formulation}

In order to achieve offset-free regulation, we adopt the standard strategies 
discussed, for example, in \cite{Pannocchia2003}. In particular, we use the following 
augmented dynamics to design an extended Kalman filter (EKF):
\begin{figure*}[t]    
\begin{tabular}{cc}
\hspace{-0.6cm}
\subfloat{
\includegraphics[scale=0.8]{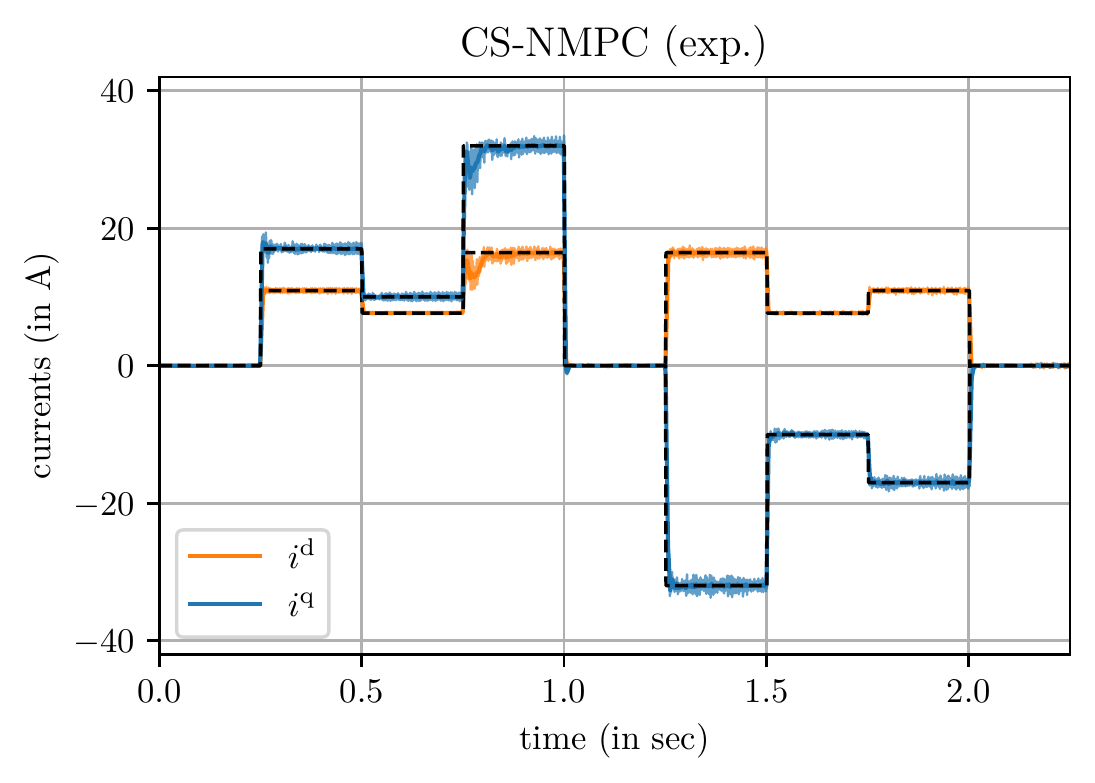}
}
&\hspace{-0.6cm}
\subfloat{
\includegraphics[scale=0.8]{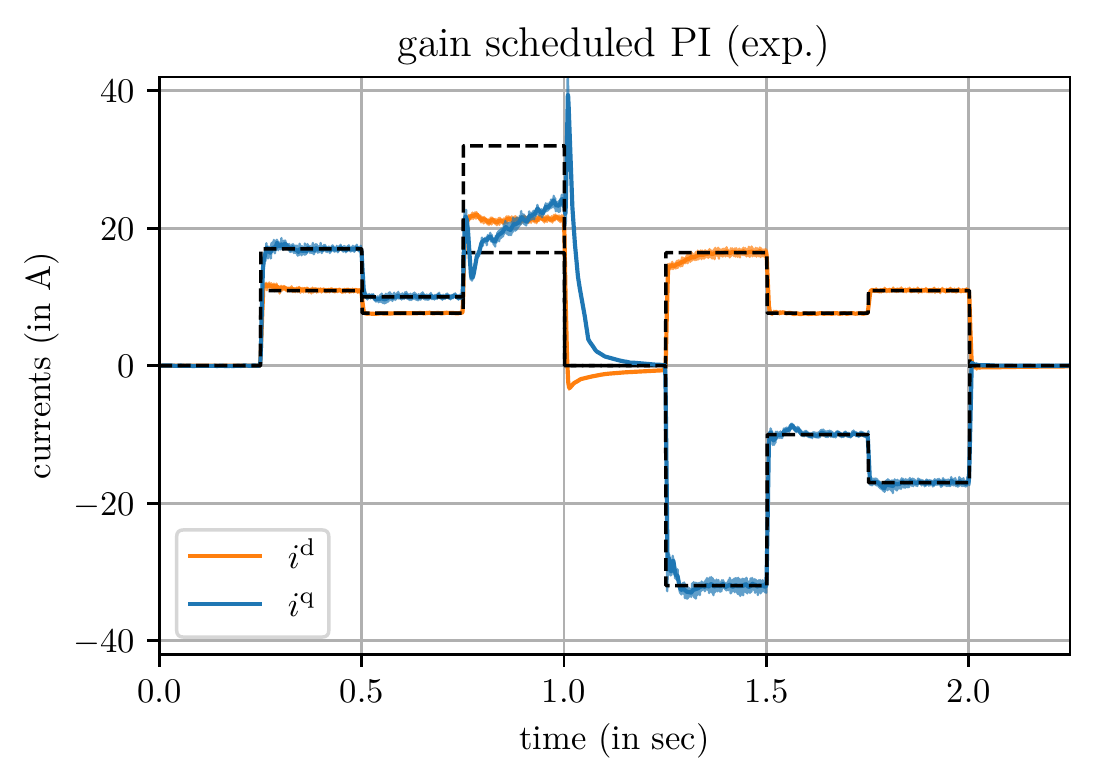}
}\end{tabular}
\caption{Current steps at $\SI{157}{\radian\per\second}$ (experiment): results obtained using the proposed CS-NMPC controller (left) and gain-scheduled PI controller (right). The CS-NMPC controller outperforms the PI controller, especially when the input constraints become active (e.g., between $t=\SI{0.75}{\second}$ and $t=\SI{1.00}{\second}$). At the same time, as it can be seen especially between $t=\SI{1.25}{\second}$ and $t=\SI{1.50}{\second}$, 
a faster transient can be achieved, even when the constraints are active only for a short time.}\label{fig:exp_TS_157_currents}
\end{figure*}

\begin{figure*}[t]    
\begin{tabular}{cc}
\hspace{-0.6cm}
\subfloat{
\includegraphics[scale=0.82]{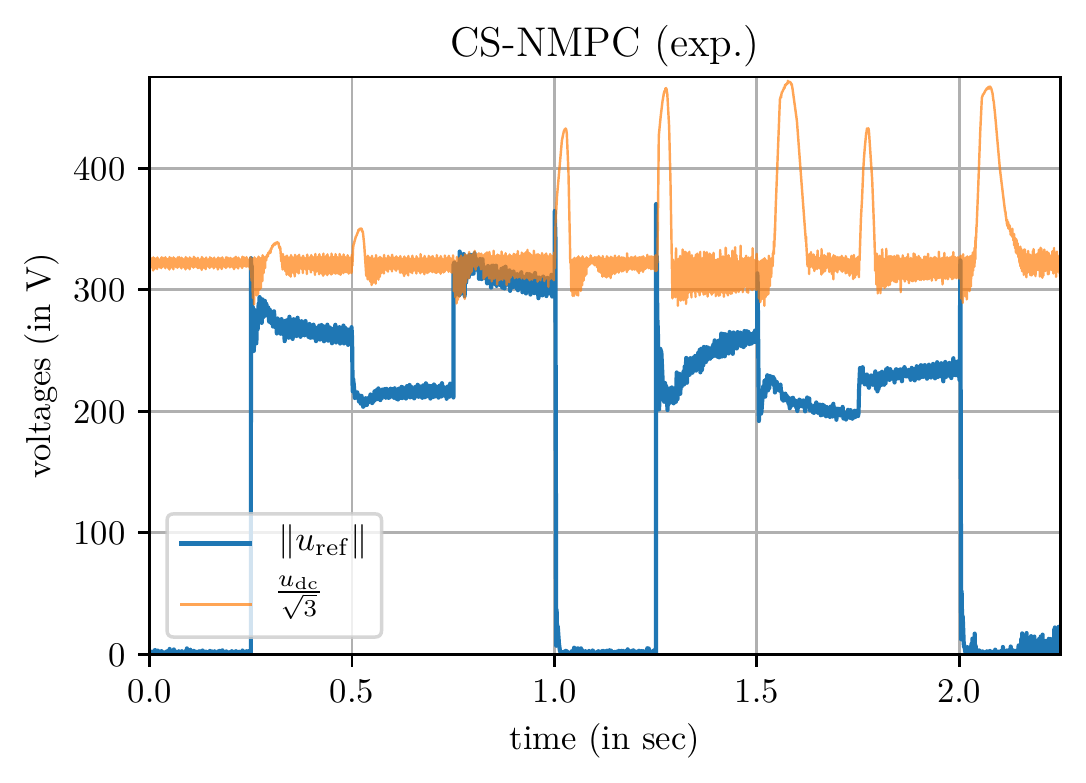}
}
&\hspace{-0.6cm}
\subfloat{
\includegraphics[scale=0.82]{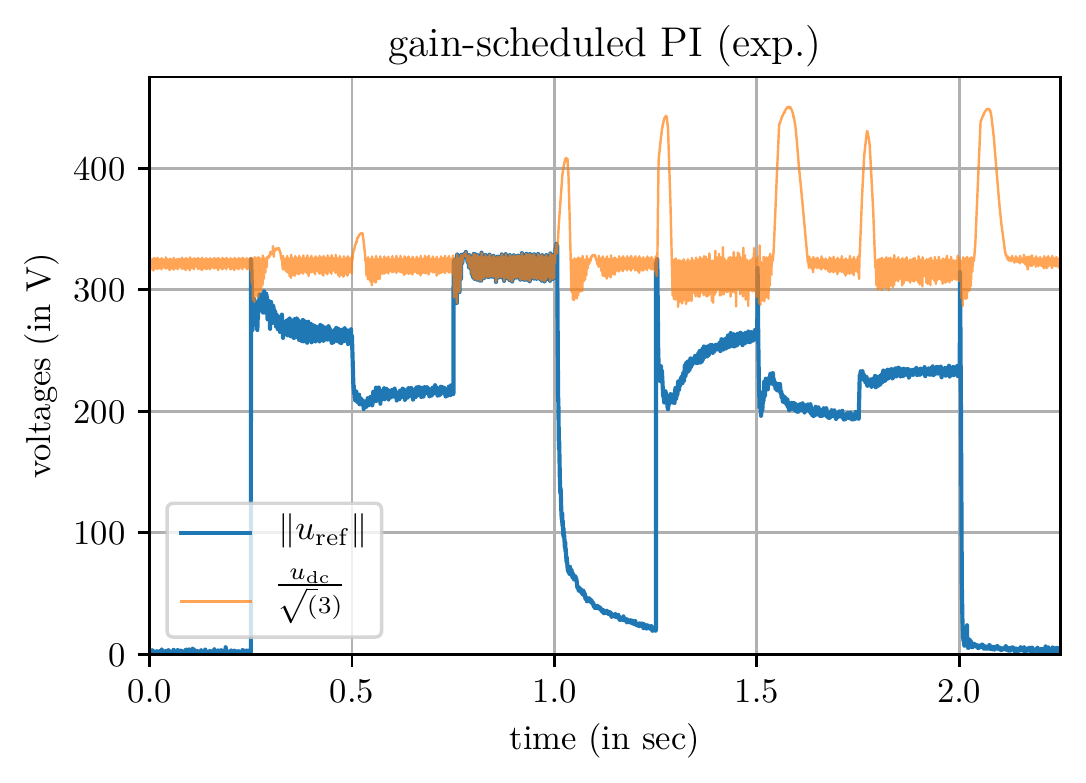}
}\end{tabular}
\caption{Current steps at $\SI{157}{\radian\per\second}$ (experiment): two-norm of measured voltage references $u_{\text{ref}}$ commanded by the two controllers and $u_{\text{dc}}$ over time. During the third current step, the PI controller saturates and does not steer the system to the desired reference. Notice that the input commanded by the PI controller remains saturated during the entire step. On the contrary, the CS-NMPC controller, after an initial saturation, steers the current to the (feasible) reference values. }\label{fig:exp_TS_157_voltages}
\end{figure*}
\begin{equation}\label{eq:DAE_EKF}
    \begin{aligned}
        &&\ddtsmall\psi &= u-R i-\omegak\J \psi + \vk, \\
        &&\ddtsmall v     &=0,\\
        &&0               &= \psi - \Psi(i), 
    \end{aligned}
\end{equation}
where the disturbance state $v$ is introduced and we assume that
pseudo-measurements $\psi_{\text{meas}}$ are available through the interpolated FEM flux maps:
\begin{equation}\label{eq:meas_fun}
    \psi_{\text{meas}} = \hat{\Psi}(i_{\text{meas}}),
\end{equation}
while current measurements $i_{\text{meas}}$ are physically carried out on the machine.
A standard EKF is designed using \eqref{eq:DAE_EKF} and \eqref{eq:meas_fun} which
uses flux measurements to estimate fluxes $\psi_{\text{e}}$ and disturbances $v_{\text{e}}$. 
Notice that the angular velocity $\omega_{\text{e}}$ is estimated externally and is considered 
as a constant-over-time parameter that is updated at every sampling time. 
In \cite[Theorem 14]{Pannocchia2015a} a streamlined version of the results from \cite{Muske2002, Pannocchia2003, Morari2012}
is presented, where under the assumptions, among others, of observability of the 
augmented dynamics \eqref{eq:DAE_EKF} and asymptotically constant disturbances 
$v$, the steady-state of the closed-loop system can be proved to be offset-free.
\section{Implementation and simulation results}\label{sec:simulations}
An RTI strategy \cite{Diehl2002a}, where a single QP of an SQP algorithm is carried out 
per sampling time, is used to solve \eqref{eq:tracking_nmpc}. In particular, the generalized 
Gauss-Newton Hessian approximation proposed in \cite{Verschueren2018} is used. In this way, 
the (positive) curvature contribution due to the convex spherical constraints on voltages 
can be exploited in order to improve the Hessian approximation used in the QP subproblems. Although 
in our experience this improves a lot the convergence of the RTI iterates on this specific problem, 
the approximate feedback law can, from time to time, be largely infeasible with respect to the 
nonlinear spherical constraints (recall that the intermediate full-step SQP iterates are feasible only 
with respect to linear constraints). Since a-posteriori projection of the control actions onto the feasible 
set can largely deteriorate the control performance, we add extra polytopic ``safety'' constraints (defined by $\hat{C}$ and $\hat{c}$ in \eqref{eq:tracking_nmpc}) 
around the spherical ones in order to ensure that the constraint violation is bounded at any successfully computed iterate.
\par 
In order to be able to meet the short sampling times required to control the electrical drive, we use a prediction horizon 
of $T_h = 3.2$ ms obtained with 2 shooting nodes ($N=2$) and we use the QP solver \texttt{qpOASES}, which is particularly suited for 
problems with short horizons \cite{Kouzoupis2018}. For both simulation and experimental results the controller is run 
at 4 kHz. 
\begin{remark}
    Notice that the discretization time ($T_h/2 = \SI{1.6}{\milli\second}$) used in the optimal control problem is not equivalent to the 
    sampling time $T_s = \SI{0.25}{\milli\second}$. This setting, which we might call ``oversampled'' NMPC, allows one to obtain 
    a longer prediction horizon without increasing the number of optimization variables. Although a theoretical analysis of this 
    strategy is well beyond the scope of this paper, we point the interested reader to \cite{Gruene2003} where it is shown that 
    fundamental properties of the feedback policy hold for the oversampled setting. Moreover, notice 
    that this setting is used in \cite{Zanelli2020} and \cite{Zanelli2020a} in order to prove asymptotic stability 
    of the combined system-optimizer dynamics.
\end{remark}
\par
We have tuned the weights in \eqref{eq:tracking_nmpc} until satisfactory closed-loop performance could be achieved 
in simulation resulting in $W=\text{blkdiag}(312.5 \cdot \mathbb{I}_2, 1 e^{-4} \cdot \mathbb{I}_2)$ and $W_N$ set 
to the corresponding LQR cost obtained with the dynamics linearized at $i=0$, $\psi=0$, $u=0$ and $\omega_{\text{m}} = 0$. Although 
the terminal cost should in general be updated together with the desired reference, a fixed terminal cost was able to provide satisfactory control performance. 
\par
In the following, we discuss simulation and experimental results 
obtained using the above described RTI strategy to solve \eqref{eq:tracking_nmpc} with \texttt{acados}.

In order to validate, first in simulation and then experimentally, the proposed approach, 
we regard a setting where the RSM is connected to a permanent synchronous machine (PMSM) which can 
be used to simulate different load conditions. The CS-NMPC controller 
has been implemented in \texttt{acados} using its Python interface 
and integrated in a \texttt{Simulink} model that makes use of a high-fidelity model of the system to be controlled including 
a model of the PMSM and of the two-level VSI described in Section \ref{sec:background}.
Moreover, we have implemented an EKF based on the augmented model \eqref{eq:DAE_EKF} using the 
implicit integrators available in $\texttt{acados}$. 
\par
We set the PMSM's controller such that it maintains a constant rotational speed and we change the torque reference 
fed to the RSM's controller to assess the tracking performance of the proposed controller. We compare
the closed-loop trajectories obtained with the ones achieved when using instead the gain-scheduled PI controller
with anti-windup presented in \cite{Article_Hackl2015_NonlinearPIcurrentcontrolofreluctancesynchronousmachines}.
The parameter tuning used in \cite{Article_Hackl2015_NonlinearPIcurrentcontrolofreluctancesynchronousmachines} was
used as baseline and we adapted the parameters until the PI controller was able to stabilize the system and achieve 
satisfactory control performance for the scenario under analysis. In particular, we
had to scale down the proportional and integral coefficients by a factor two. For the 
sake of reproducibility the entire simulation environment is made available at 
\href{https://github.com/zanellia/cs_nmpc_rsm}{\texttt{https://github.com/zanellia/cs\_nmpc\_rsm}}.
\begin{figure}[t]
	\centering
	\includegraphics[scale=0.73]{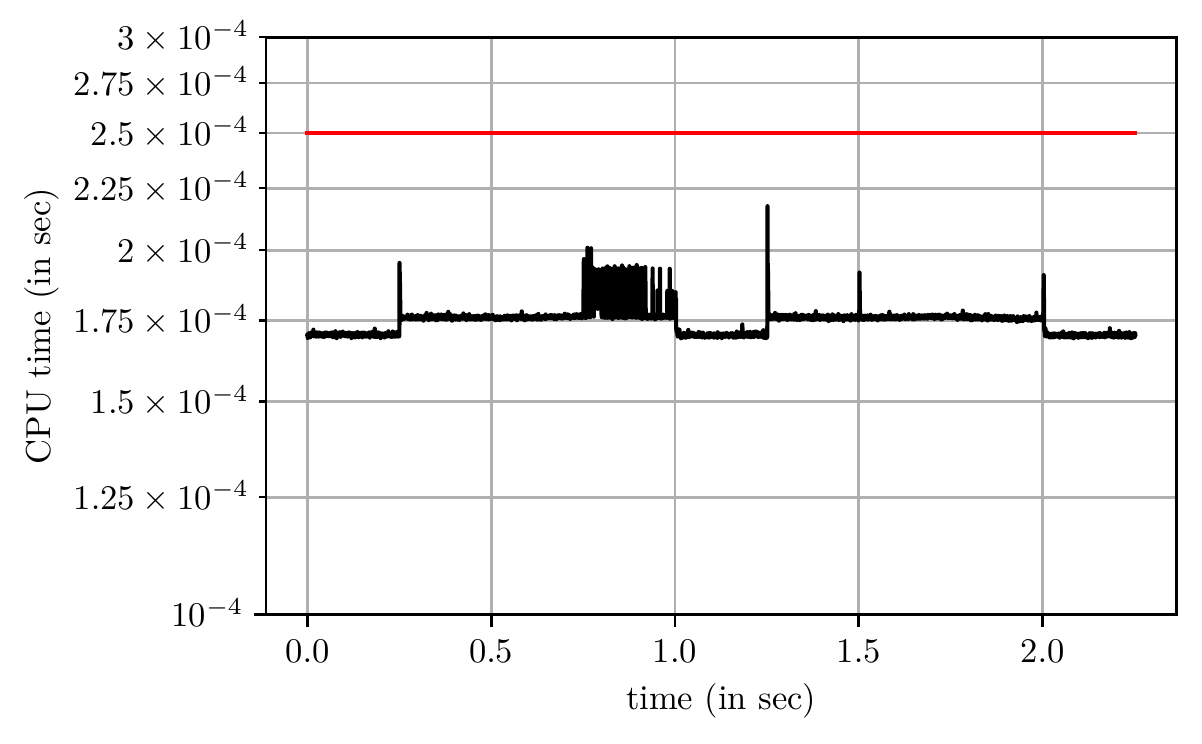}
    \caption{Current steps at $\SI{157}{\radian\per\second}$: overall control loop turnaround time (in black) obtained with the CS-NMPC controller using 
    \texttt{acados} with \texttt{qpOASES} (available computation time of \SI{250}{\micro\second} in red). About $90\%$ of the computation time 
    is due the CS-NMPC controller (together with the EKF).}\label{fig:exp_TS_157_nmpc_cpu_time}
\end{figure}
\par
In order to highlight the advantages of using a controller which can handle constraints directly, we set the reference 
speed to a value that is close to the limit value $\omega_{\text{m}}^{\star}$ computed as follows:
\begin{equation}
\begin{aligned}
    &\omega_{\text{m}}^{\star} \vcentcolon = &&\frac{1}{2}\arg \underset{\omega}{\max}  &&\omega \\
    & && \quad \quad \quad \text{s.t.} && \| R_s i_{\text{ref}} + \omega J\psi_{\text{ref}}\|_2 \leq \left(\frac{u_{\text{dc}}}{\sqrt{3}}\right).
\end{aligned}
\end{equation}
Since the optimal value is achieved at the boundaries of the feasible set, we can simply solve for $\omega$ the quadratic equation
\begin{equation}
\| R_s i_{\text{ref}} + \omega J\psi_{\text{ref}}\|^2_2 - \left(\frac{u_{\text{dc}}}{\sqrt{3}}\right)^2 = 0,
\end{equation}
such that, for the values $i_{\text{ref}} = (\SI{16.45}{}, \SI{31.99}{}) \,\,\SI{}{\ampere}$  and $\psi_{\text{ref}} = (\SI{0.819}{}, \SI{0.417}{}) \,\,\SI{}{\weber}$ associated with the 
torque value $\SI{58}{\newton\meter}$ and 
$u_{\text{dc}} = \SI{556}{\volt}$, we obtain $\omega_{\rm m}^{\star} = \SI{169.32}{\rad\per\second}$. Hence, we set $\omega_{\text{m}, \text{ref}} = \SI{157}{\rad\per\second} \approx \omega_{\text{m}, \text{nom}}$ 
for both the simulation and experimental scenarios. 
Finally, the parameters used in the simulations match the ones of the physical setup and are reported in 
Table~\ref{table:parameters}.
The current trajectories obtained with the CS-NMPC and PI controller are reported in Figure  
\ref{fig:sim_TS_157_currents} (similarly for input trajectories in Figure \ref{fig:sim_TS_157_voltages}).
It is clear from the plots that the tracking performance achieved by the CS-NMPC controller is largely superior to 
the one obtained by the PI controller, especially when the input constraints become active (e.g., between $t=\SI{0.75}{\second}$ and $t=\SI{1.00}{\second}$).
At the same time, as it can be seen from the current trajectories in Figure \ref{fig:sim_TS_157_voltages} between $t=\SI{1.25}{\second}$ and $t=\SI{1.50}{\second}$, 
a faster transient can be achieved, even when the constraints are active only for a short time.
In Appendix \ref{app:additional_plots} we report additional results obtained with a slightly increased 
reference speed $\omega_{\text{m}, \text{ref}} = \SI{165}{\rad\per\second}$. 
\par 
% Finally, in Figure \ref{fig:different_horizons}, different horizon lengths $N \in \{2,\,5,\,10,\,20\}$ are compared. 
% Although, longer horizons can achieve shorter settling time and smaller overshot, due to the limited computational 
% power available, $N=2$ is used for both simulation and experimental validation since it gives a reasonable trade-off 
% between computation times and closed-loop performance.
%\begin{figure*}[t]
%\begin{tabular}{cc}
%\subfloat[fitting error $\Psisdhat - \Psisd$]{
%% \centering
%\!\!\!\!\!\!\!
%\includegraphics[scale=0.8]{Figures/pdf/nmpc_data.pdf}}
%% \qquad
%&
%% \qquad
%%
%\subfloat[fitting error $\Psisqhat - \Psisq$]{
%% \centering
%% \includegraphics[scale=0.8]{Figures/pdf/fit_data_error_q.pdf}}
%\includegraphics[scale=0.8]{Figures/pdf/pi_data.pdf}}
%\end{tabular}
%\caption{Flux maps fitting error for grey box model.} \label{fig:fit_error}%\SI{9600}{\watt}
%\end{figure*}
\section{Experimental Results} \label{sec:experiments}

The presented NMPC scheme has been deployed on a custom-built $\SI{9.6}{\kilo\watt}$ RSM
(Courtesy of Prof.~Maarten Kamper, Stellenbosch University, South Africa) with the parameters
reported in Table~\ref{table:parameters}.
%
% \begin{equation}
%   \begin{aligned}
% 	  \Rs    &= \SI{0.4}{\ohm},         &\omegaMnom &= \SI{157.07}{\radian\per\second}, \\ 
% 	  \mMnom &= \SI{61}{\newton\meter}, &\ismax     &=  \SI{29.7}{\ampere}, \\
% 	  \usmax &= \SI{556}{\volt},               
%   \end{aligned}
%  \label{eq:Stellenbosch RSM parameters}
% \end{equation}
%
and the nonlinear flux linkage maps as depicted in Figure~\ref{fig:fit} (maps were obtained from FEM). The overall
laboratory setup is depicted in Figure~\ref{fig:lab_setup} and comprises the \texttt{dSPACE} real-time system with processor
board DS1007 and various extensions and I/O boards, two \SI{22}{\kilo\watt} SEW inverters in back-to-back
configuration sharing a common DC-link. Moreover, it comprises the HOST-PC running \texttt{MATLAB}/\texttt{Simulink} 
with RCPHIL R2017 and \texttt{dSPACE
ControlDesk} 6.1p4 for rapid-prototyping, data acquisition and evaluation, the custom-built
\SI{9.6}{\kilo\watt} RSM  as device under test and a \SI{14.5}{\kilo\watt} SEW PMSM  as load machine. The
DR2212 torque sensor allows to measure the mechanical torque too, but it was not used during the experiments.
The CS-NMPC controller based on the formulation described in Section \ref{sec:simulations}
and implemented using the \texttt{acados} framework has been deployed on the 
\texttt{dSPACE} unit connected to the physical RSM. 
\par
In order to validate the control performance of the proposed CS-NMPC controller we reproduced the scenario 
used for the simulation reported in Section \ref{sec:simulations}, i.e., we 
used the PMSM to maintain the nominal rotational speed 
of the rotor ($\SI{157}{\radian\per\second}$) and different torque references have been fed to the RSM controllers under 
analysis.
\par 
The closed-loop trajectories for the conducted experiments are reported in Figure \ref{fig:exp_TS_157_currents}-\ref{fig:exp_TS_157_voltages}.
Similarly to the results obtained in simulation (see Figures \ref{fig:sim_TS_157_currents} and \ref{fig:sim_TS_157_voltages}), the proposed 
CS-NMPC controller achieves better tracking performance than the gain-scheduled PI controller, 
in particular when the voltage constraint becomes active (especially between $t=\SI{0.75}{\second}$ and $t=\SI{1.00}{\second}$). Notice that there is a non negligible 
discrepancy between simulation and experimental results potentially due to model mismatches. Among other possible causes of discrepancies, we mention the fact 
that the modelled flux maps might differ from the real ones. Moreover, we observe from Figure \ref{fig:exp_TS_157_voltages} that the DC-link voltage fluctuates around its nominal value. 
In fact, in the presence of a sudden change in the torque reference, the voltage of the DC-link capacitor 
can drop if the recharging rate is slower than the discharging rate (behavior not modelled in simulation). Although this behavior 
is not accounted for in the dynamics of the system used to design the controllers under analysis, in both controllers we exploit the measured DC-link voltage in order to adjust the 
feasible set. 
Notice that the $u_{\text{ref}}$ computed by the CS-NMPC controller becomes sometimes infeasible. This infeasibility is consistent with the fact that the quadratic constraint can be violated during transient and will only be satisfied at the steady. However, due to the safety polytopic constraints introduced in \eqref{eq:tracking_nmpc}, we can guarantee that the computed solution will not violate the outer voltage hexagon.
At the same time, whenever a voltage reference that lies outside of the circular feasible set is fed to the modulator, a projection onto the disk of radius $\frac{u_{\text{dc}}}{\sqrt{3}}$ is carried out.

\section{Conclusions}
In this paper we present simulation and experimental results
obtained with a CS-NMPC torque controller for RSMs. As opposed to most 
successful implementations present in the literature, that use instead 
FS-MPC/NMPC, we show the effectiveness and real-time feasibility 
of the continuous control set approach. In particular, we show that, using the software implementation 
of the \textit{real-time iteration method} for NMPC available in the software 
package \texttt{acados}, it is possible to deploy the proposed controller on 
embedded hardware and to meet the challenging sampling times typically required  to 
control electrical drives. 
We discuss implementation details and report on simulation as well as experimental results 
which show that the proposed approach can 
largely outperform state-of-art control methods especially when the input constraints become active.
\par
Future research will involve the investigation of novel numerical methods, e.g., 
the real-time first-order methods proposed in \cite{Zanelli2019a}, to speed up the computation times, which  
are currently still rather long and neither allow for extensions of the optimal control formulations
(e.g longer horizons, state or input spaces of higher dimension, etc) nor for deployment on hardware with 
lower computational power.
\begin{table}
    \centering
\def\arraystretch{1.1}%  1 is the default, change whatever you need
\begin{tabular}{lllll}
      \hline
      par. & value & par. & value &  \\
      \hline
      $\Rs$    & $\SI{0.4}{\ohm}$         &$\omegaMnom$ & $\SI{157.07}{\radian\per\second}$ \\ 
	  $\mMnom$ & $\SI{61}{\newton\meter}$ &$\ismax$     &  $\SI{29.7}{\ampere}$ \\
      $\usmax$ & $\SI{556}{\volt}$  & $-$& &     \\        
      \hline
\end{tabular}
\caption{Parameters of physical setup (used in simulation too).}\label{table:parameters}
\end{table}

% if have a single appendix:
%\appendix[Proof of the Zonklar Equations]
% or
%\appendix  % for no appendix heading
% do not use \section anymore after \appendix, only \section*
% is possibly needed

% use appendices with more than one appendix
% then use \section to start each appendix
% you must declare a \section before using any
% \subsection or using \label (\appendices by itself
% starts a section numbered zero.)
%

% \section*{Acknowledgment}
% The authors would like to thank...

% Can use something like this to put references on a page
% by themselves when using endfloat and the captionsoff option.
\ifCLASSOPTIONcaptionsoff
  \newpage
\fi

% trigger a \newpage just before the given reference
% number - used to balance the columns on the last page
% adjust value as needed - may need to be readjusted if
% the document is modified later
%\IEEEtriggeratref{8}
% The "triggered" command can be changed if desired:
%\IEEEtriggercmd{\enlargethispage{-5in}}

% references section

% can use a bibliography generated by BibTeX as a .bbl file
% BibTeX documentation can be easily obtained at:
% http://mirror.ctan.org/biblio/bibtex/contrib/doc/
% The IEEEtran BibTeX style support page is at:
% http://www.michaelshell.org/tex/ieeetran/bibtex/
%\bibliographystyle{IEEEtran}
% argument is your BibTeX string definitions and bibliography database(s)
%\bibliography{IEEEabrv,../bib/paper}
%
% <OR> manually copy in the resultant .bbl file
% set second argument of \begin to the number of references
% (used to reserve space for the reference number labels box)
\bibliographystyle{plain}
% \bibliography{/include/MyBIB_allinone, syscop}
\bibliography{syscop}
% \bibliography{syscop} 

% biography section
% 
% If you have an EPS/PDF photo (graphicx package needed) extra braces are
% needed around the contents of the optional argument to biography to prevent
% the LaTeX parser from getting confused when it sees the complicated
% \includegraphics command within an optional argument. (You could create
% your own custom macro containing the \includegraphics command to make things
% simpler here.)
%\begin{IEEEbiography}[{\includegraphics[width=1in,height=1.25in,clip,keepaspectratio]{mshell}}]{Michael Shell}
% or if you just want to reserve a space for a photo:

% \begin{comment}
% \par
% \vspace{-1.0cm}
% \par
\begin{IEEEbiography}[{\includegraphics[width=1in,height=1.25in,clip,keepaspectratio]{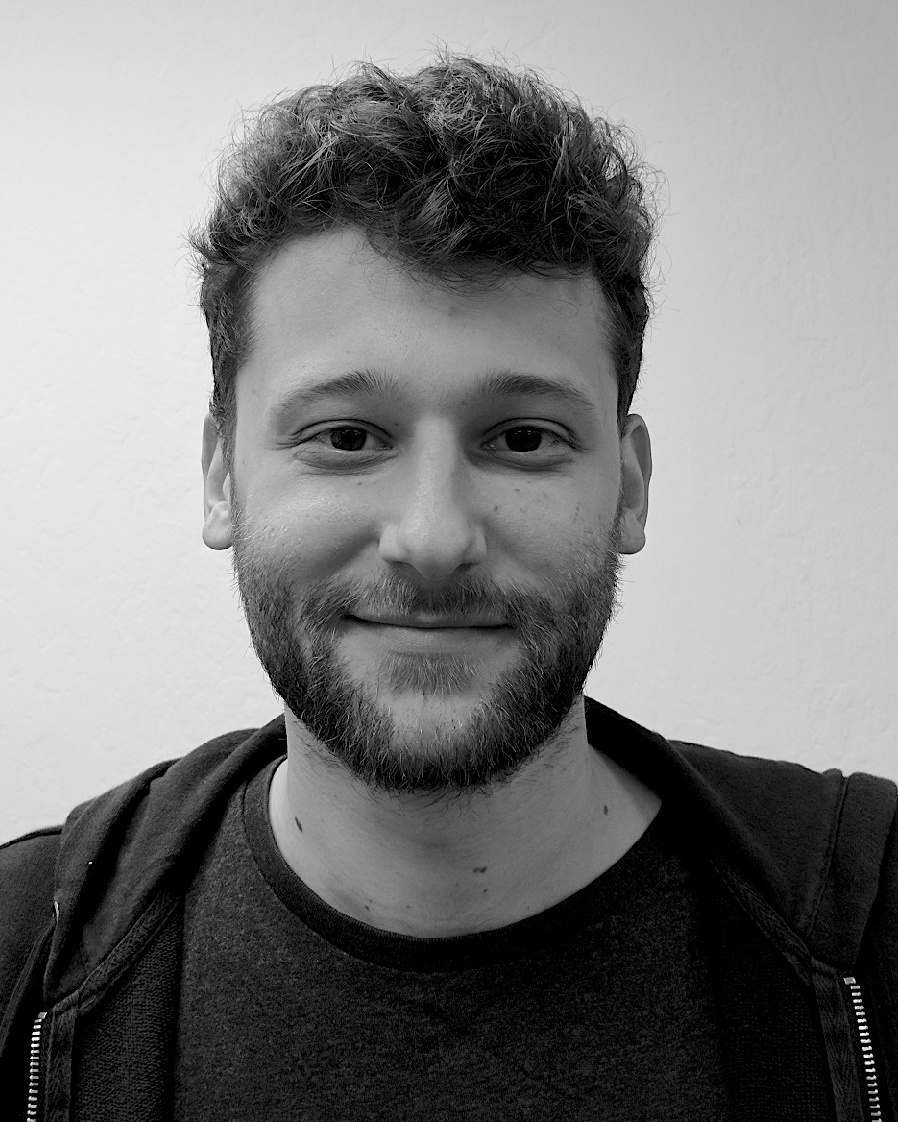}}]{Andrea Zanelli}
received the B.Sc degree in Automation Engineering from Politecnico di Milano and the M.Sc in Robotics, Systems and Control from ETH 
Zurich in 2012 and 2015, respectively. Since 2015, he is a Ph.D student at the System, Control and Optimization Laboratory at the University 
of Freiburg, Germany. His research work focuses on the development and software implementation of efficient numerical methods for embedded optimization and nonlinear model predictive control with numerical and system theoretic guarantees.
\end{IEEEbiography}
% \par
% \vspace{-1.0cm}
% \par
\begin{IEEEbiography}[{\includegraphics[width=1in,height=1.25in,clip,keepaspectratio]{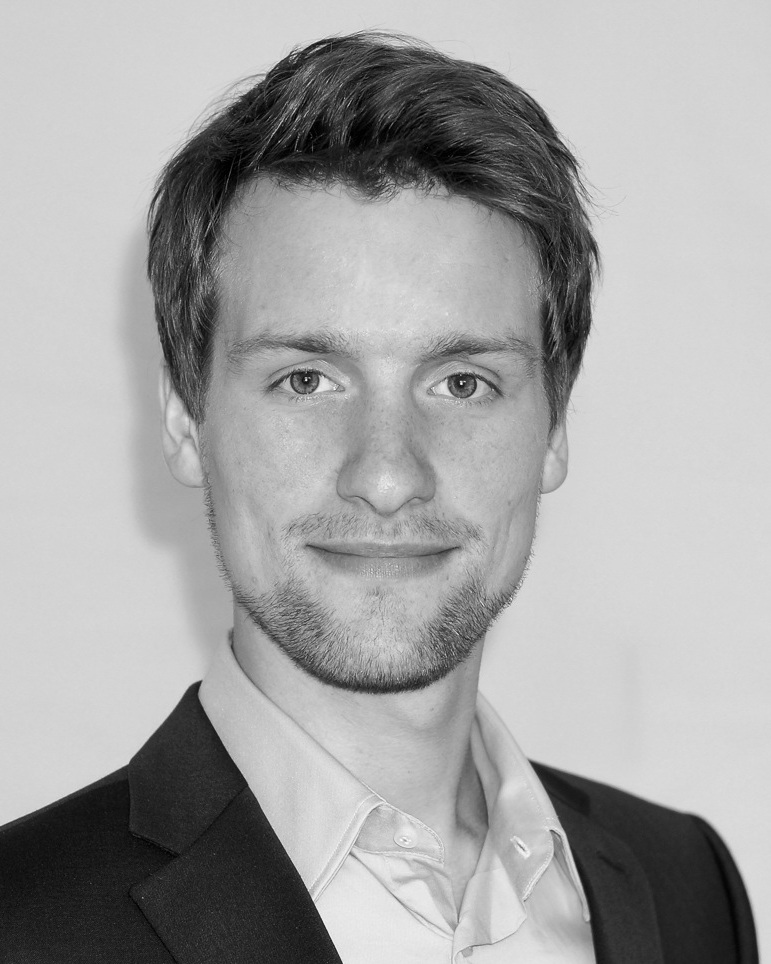}}]
{Julian Kullick} received the B.Sc. and M.Sc. degrees in electrical engineering from Technical University of Munich (TUM), Munich, 
Germany in 2012 and 2015
respectively. He is currently working toward the Ph.D. degree in electrical engineering with TUM. From 2015 to 2018, in
the course of the ``Geothermal-Alliance Bavaria (GAB)'' project, he was a Research Associate with the research group
``Control of Renewable Energy Systems (CRES)'' at TUM.  Since 2019 he is a Research Associate with the ``Laboratory for
Mechatronic and Renewable Energy Systems (LMRES)'', Munich University of Applied Sciences, Munich, Germany. His research
interests include nonlinear modeling, efficient operation and sensorless control of electric drives.
\end{IEEEbiography}
\begin{IEEEbiography}[{\includegraphics[width=1in,height=1.25in,clip,keepaspectratio]{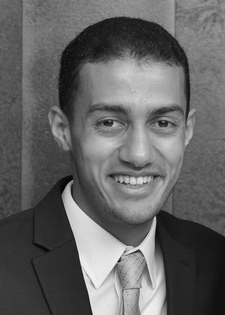}}]{Hisham M. Eldeeb} received his B.Sc. (honors) and  M.Sc.  in  electrical  engineering  in  2011 and 2014,   respectively,   from   the   Faculty   of   Engineering,  Alexandria  University,  Egypt.  From 2012  to  2015,  he  worked  as  a  research  associate  at  Qatar  University  in  Qatar;  
    aiming  at  extending  the  penetration  of  inverter-based  distributed-generation plants in the Qatari-Network. 
From September 2015, he was selected as one of the 14 Marie-Curie Ph.D. candidates, hired  on  the Horizon  H2020  project  "Airborne  Wind  Energy System Control and Optimization" (AWESCO), and earned his Ph.D degree at the Technical University of Munich (TUM) in Germany. His Ph.D topic was related to multi-phase drives and the development of fault-tolerant control strategies. He is currently working as an electrical drive specialist and development engineer for e-vehicles at  IAV  GmbH.  His  research  interests  are  grid-connected  converters, power electronics, and multi-phase drives.
\end{IEEEbiography}
\begin{IEEEbiography}[{\includegraphics[width=1in,height=1.25in,clip,keepaspectratio]{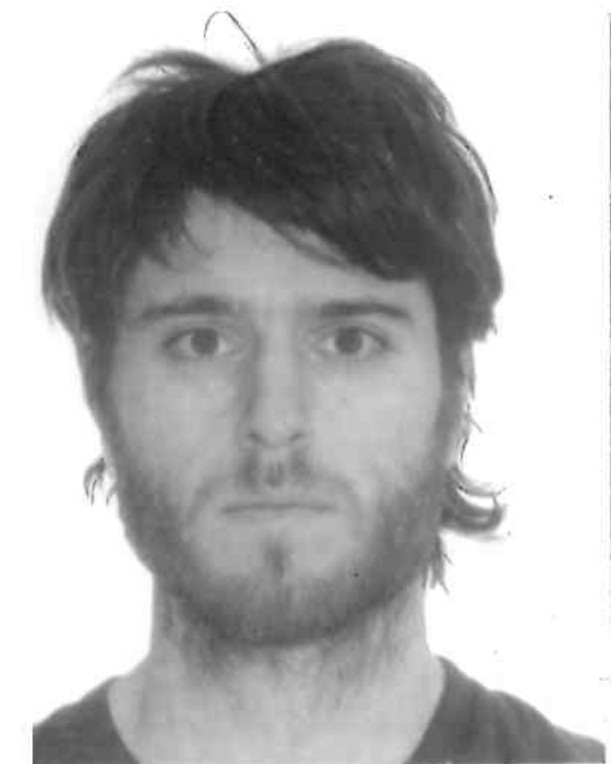}}]
{Gianluca Frison}
received the B.Sc. in Information Engineering in 2009 and the M.Sc. in Automation
Engineering in 2012, both from University of Padua. In 2012 he also received the M.Sc. in Mathematical
Modeling and Computation, followed in 2016 by the Ph.D, both from Technical University of Denmark.
Since January 2016 he has been Post-Doc, alternatively at University of Freiburg, at Technical University
of Denmark, and again at University of Freiburg. His current research interests include high-performance
linear algebra for embedded optimization and fast numerical methods for model predictive control. He is the
main developer of the software packages \texttt{BLASFEO} and \texttt{HPMPC}/\texttt{HPIPM}.
\end{IEEEbiography}
\begin{IEEEbiography}[{\includegraphics[width=1in,height=1.25in,clip,keepaspectratio]{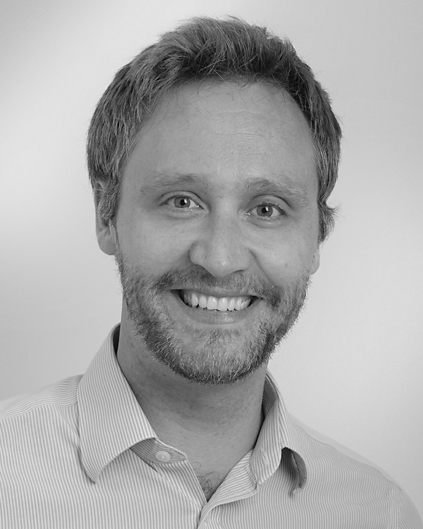}}]{Christoph M. Hackl}
(M'12-SM'16) studied Electrical Engineering (with focus on controls and mechatronics) at Technical University of Munich (TUM), Germany and
University of Wisconsin-Madison, USA, and received the B.Sc., Dipl.-Ing., and Dr.-Ing. (Ph.D.) degrees in Electrical
Engineering in 2003, 2004 and 2012, respectively, from TUM. Since 2004, he has been teaching
electrical drives, power electronics, and mechatronic \& renewable
energy systems. Since 2014, he has been the head of the research group ``Control of Renewable Energy Systems (CRES)'' at
TUM.  In 2018, he became a Professor for Electrical Machines and Drives and the head of the ``Laboratory for Mechatronic
and Renewable Energy Systems (LMRES)'' at the Munich University of Applied Sciences (MUAS), Germany. 
His research interests include nonlinear, adaptive and optimal control of electrical, mechatronic and renewable energy
systems.
\end{IEEEbiography}
\begin{IEEEbiography}[{\includegraphics[width=1in,height=1.25in,clip,keepaspectratio]{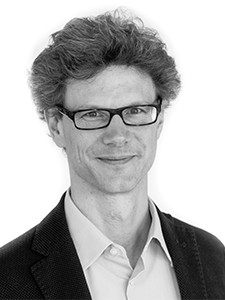}}]{Moritz Diehl}
graduated with a Diploma in physics and mathematics at Heidelberg University,
Heidelberg, Germany and Cambridge University, Cambridge, U.K. in 1999, and received his
Ph.D. degree from Heidelberg University in 2001.
From 2006 to 2013, he was a Professor with the Department of Electrical Engineering, KU Leuven,
Leuven, Belgium. In 2013, he moved to the University of Freiburg, Freiburg, Germany, where he is
currently the Head of the Systems Control and Optimization Laboratory, Department of Microsystems
Engineering, and the Department of Mathematics
\end{IEEEbiography}
% \end{comment}

% You can push biographies down or up by placing
% a \vfill before or after them. The appropriate
% use of \vfill depends on what kind of text is
% on the last page and whether or not the columns
% are being equalized.
\vspace{0.5cm}
\appendix[Additional simulation and experimental results]\label{app:additional_plots}

\begin{figure*}[h]    
\begin{tabular}{cc}
\hspace{-0.6cm}
\subfloat{
    \includegraphics[scale=0.8]{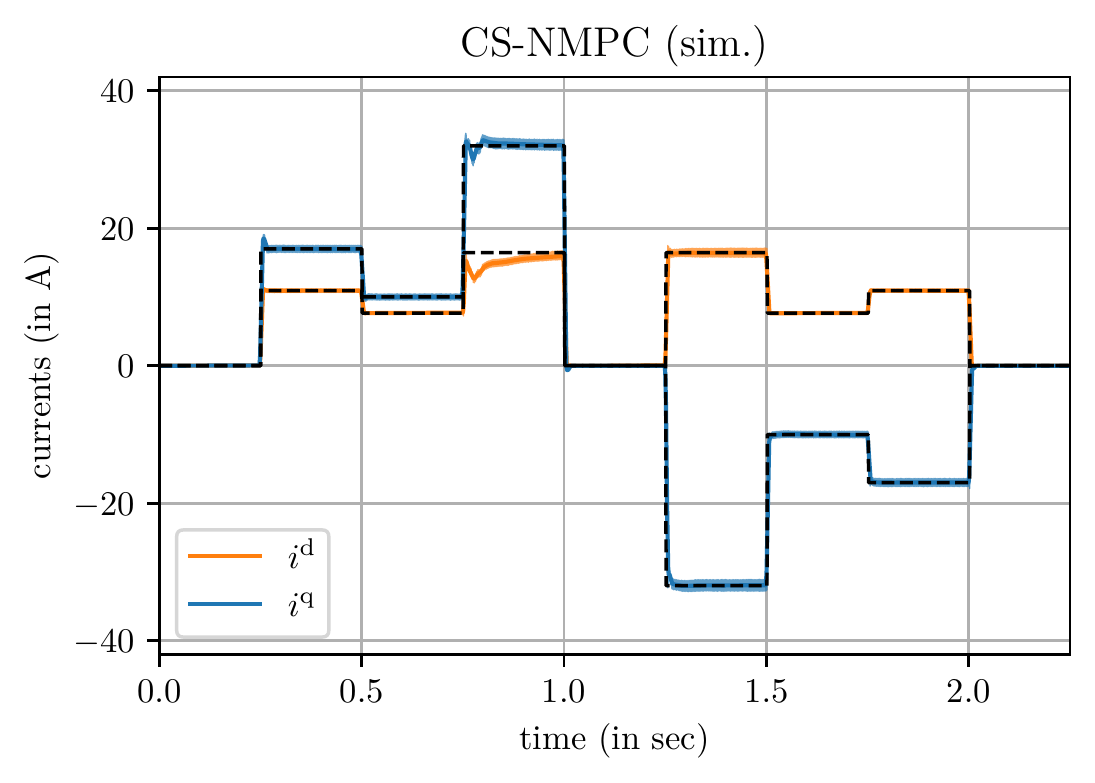}
}
&\hspace{-0.6cm}
\subfloat{
    \includegraphics[scale=0.8]{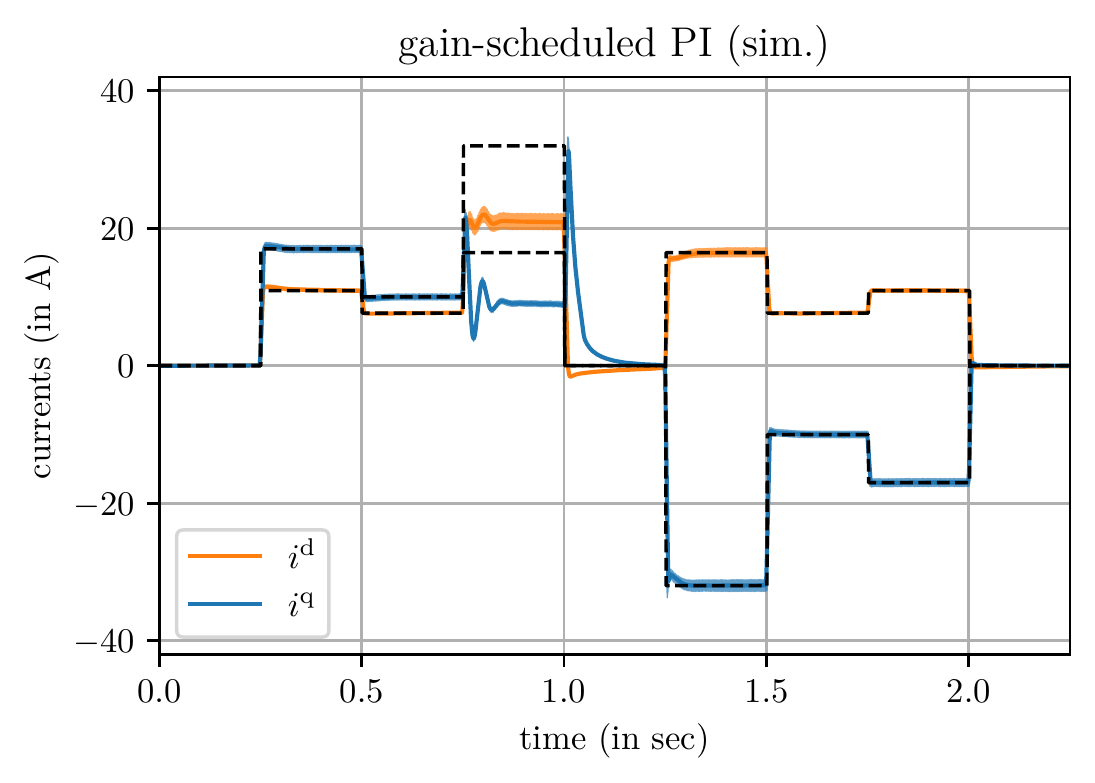}
}
\end{tabular}
\caption{Current steps at $\SI{165}{\radian\per\second}$ (simulation): closed-loop trajectories obtained using CS-NMPC (left) and gain-scheduled PI controller (right).}\label{fig:sim_TS_165_currents_PI}
\end{figure*}

\begin{figure*}[t]    
\begin{tabular}{cc}
\hspace{-0.6cm}
\subfloat{
\includegraphics[scale=0.8]{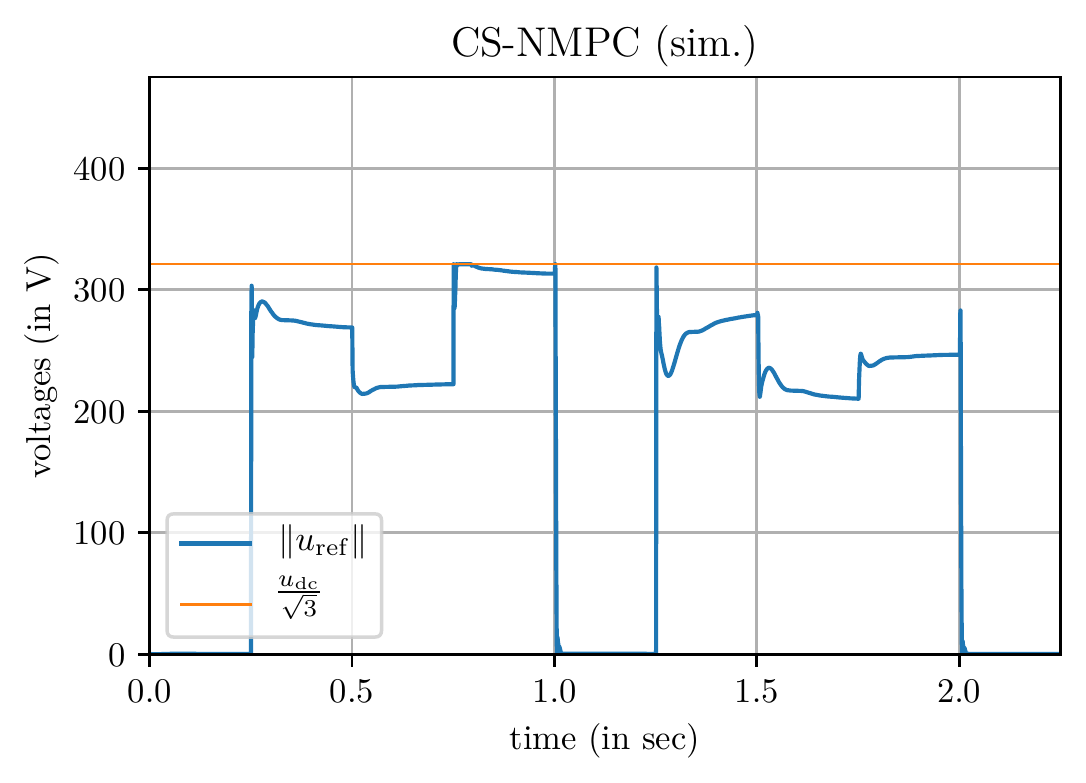}
}
&\hspace{-0.6cm}
\subfloat{
\includegraphics[scale=0.8]{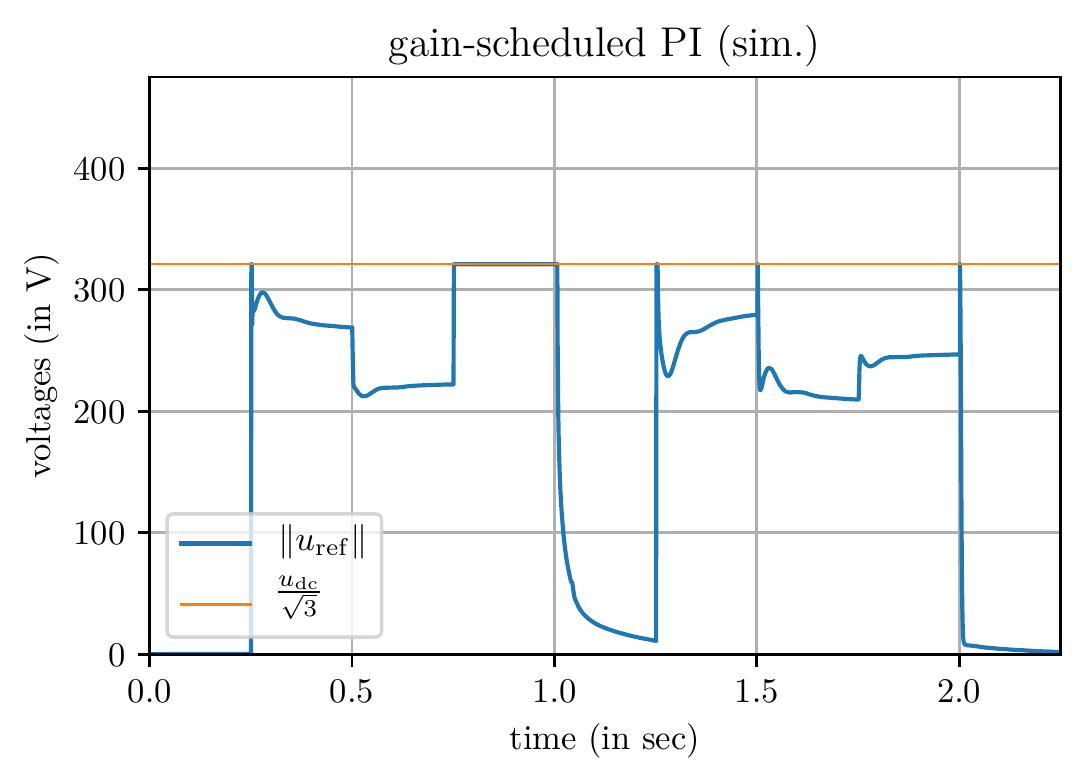}
}\end{tabular}
\caption{Current steps at $\SI{165}{\radian\per\second}$ (simulation): two-norm of voltage references $u_{\text{ref}}$ commanded by the two controllers and $u_{\text{dc}}$ over time. During the third current step, the PI controller saturates and does not steer the system to the desired reference.}\label{fig:exp_TS_165_voltages}
\end{figure*}

\begin{figure*}[t]    
\begin{tabular}{cc}
\hspace{-0.6cm}
\subfloat{
    \includegraphics[scale=0.8]{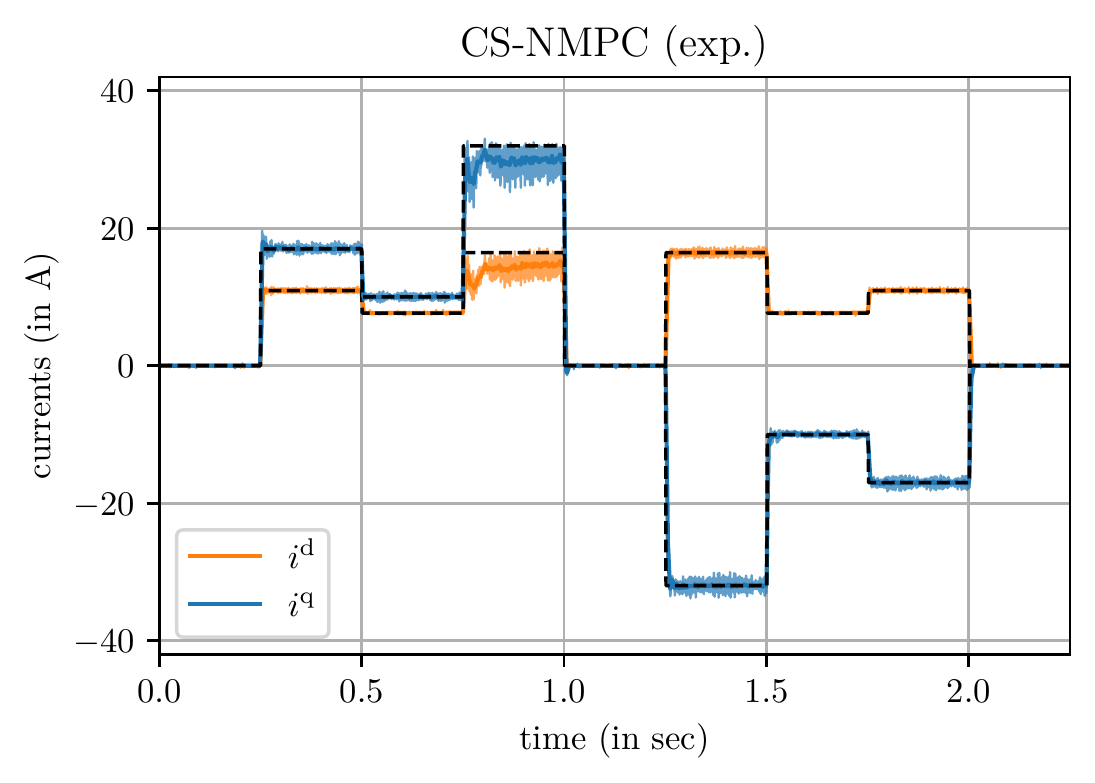}
}
&\hspace{-0.6cm}
\subfloat{
    \includegraphics[scale=0.8]{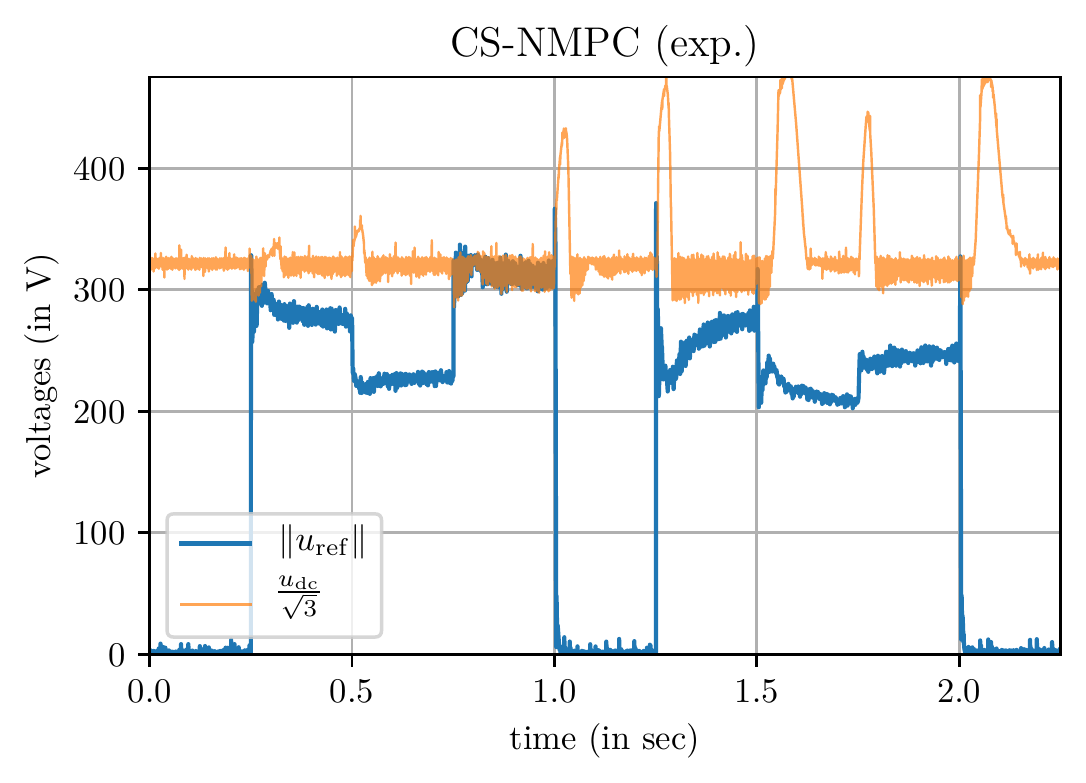}
}\end{tabular}
\caption{Current steps at $\SI{165}{\radian\per\second}$ (experiment): closed-loop current trajectories obtained using the two controllers under analysis. Due to the strong effect of input saturation 
observed in simulation, for this value of the reference speed it was not possible to run the experiment with the PI controller.}\label{fig:exp_TS_165_currents_NMPC}
\end{figure*}

In the following, we report additional simulation and experimental (for CS-NMPC only)
results carried out at a reference speed of $\SI{165}{\rad\per\second}$. In this 
scenario, the impact of the input constraint is even stronger than for the one 
used in Sections \ref{sec:simulations} and \ref{sec:experiments}. Since the 
results showed potentially damaging behavior when controlling the RSM with the gain-scheduled PI, 
we ran the corresponding experiments only with CS-NMPC. The simulation and 
experimental results are reported in Figures \ref{fig:sim_TS_165_currents_PI},
\ref{fig:exp_TS_165_voltages} and \ref{fig:exp_TS_165_currents_NMPC}.
These additional results confirm the observation made for the scenario reported 
in Sections~\ref{sec:simulations} and \ref{sec:experiments}.
\vspace{2cm}
% Notice that, unlike in simulation, in the experimental validation the CS-NMPC controller is not able to steer the system to the desired 
% steady-state. This is likely due to flux model mismatch 

% \begin{figure}[h]    
%     \includegraphics[scale=0.82]{Figures/pdf/NMPC_165_RAD_S_58_NM_BIG_sim_revision.pdf}
% \caption{Current steps at $\SI{165}{\radian\per\second}$ (simulation): closed-loop trajectories obtained using the CS-NMPC controller.}\label{fig:sim_TS_165_currents_NMPC}
% \end{figure}

% \begin{figure}[h]    
%     \includegraphics[scale=0.82]{Figures/pdf/PI_165_RAD_S_58_NM_BIG_sim_revision.pdf}
% \caption{Current steps at $\SI{165}{\radian\per\second}$ (simulation): closed-loop trajectories obtained using the gain-scheduled PI controller.}\label{fig:sim_TS_165_currents_PI}
% \end{figure}

% \begin{figure}[h]    
% \includegraphics[scale=0.8]{Figures/pdf/NMPC_165_RAD_S_58_NM_BIG_u_norm_revision.pdf}
% \caption{Current steps at $\SI{165}{\radian\per\second}$ (simulation): two-norm of voltage references $u_{\text{ref}}$ commanded by the two controllers and maximum two-norm $u_{\text{dc}}$ over time. During the third current step, the PI controller saturates and does not steer the system to the desired reference. Notice that the input commanded by the PI controller remains saturated during the entire step. On the contrary, the CS-NMPC controller, after an initial saturation, steers the current to the (feasible) reference values. }\label{fig:exp_TS_165_voltages_NMPC}
% \end{figure}

% \begin{figure}[h]    
% \includegraphics[scale=0.8]{Figures/pdf/PI_165_RAD_S_58_NM_BIG_u_norm_revision.pdf}
% \caption{Current steps at $\SI{165}{\radian\per\second}$ (simulation): two-norm of voltage references $u_{\text{ref}}$ commanded by the two controllers and maximum two-norm $u_{\text{dc}}$ over time. During the third current step, the PI controller saturates and does not steer the system to the desired reference. Notice that the input commanded by the PI controller remains saturated during the entire step. On the contrary, the CS-NMPC controller, after an initial saturation, steers the current to the (feasible) reference values. }\label{fig:exp_TS_165_voltages_PI}
% \end{figure}

\end{document}

%% file: include/MyPackages_General_EN.tex
\usepackage[english]{babel}
\def\germanlanguage{german}

\usepackage[latin1]{inputenc}
\usepackage[T1]{fontenc}

\usepackage{pifont}		%	allows the use of e.g. \ding{172} = encircled 1 or \checkmark
\usepackage[active]{srcltx}
\usepackage{eurosym}

\usepackage{subdepth} % -> all symbols with subscript are on one line)
\usepackage{siunitx}
\DeclareSIUnit{\var}{var}
\DeclareSIUnit{\rad}{rad}
\DeclareSIUnit{\umdrehung}{U}
\usepackage{lscape}		%	allow to use landscape environments
\usepackage{framed}		%	allows to insert frames (boxes) around text, etc.. BUT GIVES ERRORS concerning \dimen ????
\usepackage{longtable}
\usepackage{booktabs}	%	nice tables with toprule, midrule, bottomrule - CONFLICT WHEN USING IN BEAMER!
\usepackage{fancyhdr}
\usepackage{setspace}
\usepackage{enumerate} 
\usepackage{multienum}
\usepackage{xcolor}
\usepackage{comment}
\usepackage[ngerman, num]{isodate}	% for using date
\usepackage{subdepth} % -> all symbols with subscript are on one line)

\usepackage{trfsigns}	% add nice symbols for transformations e.g. \laplace = empty circle + line + filled circle (time to Laplace domain) or \Laplace = filled circle + line + empty circle (backwards, from Laplace to time domain)
\usepackage{mathrsfs} % math fonts for systems (plant), controllers, sensors, ... in block diagrams			
\usepackage{mathdots}	%	for inverse diagonal dots
\usepackage{cancel}
\usepackage{amsmath}
\usepackage{amssymb}
\usepackage{amsfonts}
\usepackage{stmaryrd}	%	for use of \llbracket and \rrbracket
\usepackage{amsthm}
\usepackage{thmtools}

\usepackage[font=footnotesize,format=hang]{caption}
\usepackage{pdfpages}
\usepackage{float}
\usepackage{subfig}
\usepackage{graphicx}
\usepackage{psfrag}
\usepackage[]{silence} % 
\usepackage[process=auto,crop=pdfcrop,cleanup={.tex,.dvi,.ps,.pdf,.snm,.out,.bbl,.nav,.auxlock,.toc}]{pstool}
\usepackage{makeidx}
\usepackage{lastpage}
\usepackage{algorithmicx}
\usepackage{listings}

\usepackage[sort&compress,numbers]{natbib}

%% file: include/MyPackages_TikZ_externalize.tex
\usepackage{tikz}
\newlength\figurewidth
\newlength\figureheight
\usepackage{pgfplots}
\pgfplotsset{compat=newest}
\usepgfplotslibrary{fillbetween}
\pgfkeys{/pgf/number format/.cd, set thousands separator={}}
\usepgfplotslibrary{external} 
\tikzset{external/up to date check={md5}}
\usepackage{tikz-timing}

\usepackage[european]{circuitikz} %
\usetikzlibrary{circuits.ee.IEC}

\usetikzlibrary{fit}
\usetikzlibrary{automata}
\usetikzlibrary{arrows}
\usetikzlibrary{calc}
\usetikzlibrary{intersections}
\usetikzlibrary{shadows}
\usetikzlibrary{backgrounds}
\usetikzlibrary{shapes}
\usetikzlibrary{positioning}
\usetikzlibrary{trees}
\usetikzlibrary{datavisualization}
\usetikzlibrary{mindmap}

  \tikzset{
    invisible/.style={opacity=0},
    visible on/.style={alt={#1{}{invisible}}},
    alt/.code args={<#1>#2#3}{%
      \alt<#1>{\pgfkeysalso{#2}}{\pgfkeysalso{#3}} % \pgfkeysalso doesn't change the path
    },
  }

\tikzset{set voltage source graphic = voltage source IEC graphic}
\tikzset{voltage source IEC graphic/.style={circuit symbol lines, circuit symbol size = width 3 height 3, shape = generic circle IEC, fill = white,
    /pgf/generic circle IEC/before background={
				\draw[-] (0, -1pt) -- (0, 1pt);
    }
}}

\tikzstyle{sum} = [draw, circle, inner sep = 0pt, minimum width=0.2cm, fill = white]

\definecolor{matlabgreen}{rgb}{0, 0.5, 0}
\definecolor{matlabmagenta}{rgb}{0.75, 0, 0.75}

\tikzset{%
  voltage info/.is family,
  voltage info/vertical offset/.initial = 1,%.5,
  voltage info/arrow direction/.initial = ->,
  voltage info/horizontal offset/.initial = 5, %10,
  voltage info/label position/.initial = above,
  voltage info/label rotate/.initial = 0,
  voltage info/label/.initial = {}
}

\tikzset{%
  voltage arrow/.style = {%
    /utils/exec = {%
					\pgfsetarrowoptions{direction ee}{%
        .4*\the\tikzcircuitssizeunit+.3*\the\pgflinewidth}},
    > = direction ee},
  voltage info sloped/.style = {% 
    append after command = {%
      \bgroup
        [voltage info/.cd,#1]
        [current point is local=true]
        [every voltage info/.try]
        [shift={($(\tikzlastnode.north)+(0,\pgfkeysvalueof{/tikz/voltage info/vertical offset}\tikzcircuitssizeunit)$)}]
        [voltage arrow,\pgfkeysvalueof{/tikz/voltage info/arrow direction}]
        (-\pgfkeysvalueof{/tikz/voltage info/horizontal offset}\tikzcircuitssizeunit,0)
        edge[line to] 
        node[\pgfkeysvalueof{/tikz/voltage info/label
          position},transform shape,rotate = \pgfkeysvalueof{/tikz/voltage
          info/label rotate}] {\pgfkeysvalueof{/tikz/voltage info/label}}
        (\pgfkeysvalueof{/tikz/voltage info/horizontal offset}\tikzcircuitssizeunit,0)
      \egroup
    }
  },
  voltage info' sloped/.style = {% 
    append after command = {%
      \bgroup
        [voltage info/.cd,label position = below,#1]
        [current point is local=true]
        [every voltage info/.try]
        [shift={($(\tikzlastnode.south)+(0,-\pgfkeysvalueof{/tikz/voltage info/vertical offset}\tikzcircuitssizeunit)$)}]
        [voltage arrow,\pgfkeysvalueof{/tikz/voltage info/arrow direction}]
        (-\pgfkeysvalueof{/tikz/voltage info/horizontal offset}\tikzcircuitssizeunit,0)
        edge[line to] 
        node[\pgfkeysvalueof{/tikz/voltage info/label
          position},transform shape,rotate = \pgfkeysvalueof{/tikz/voltage
          info/label rotate}] {\pgfkeysvalueof{/tikz/voltage info/label}}
        (\pgfkeysvalueof{/tikz/voltage info/horizontal offset}\tikzcircuitssizeunit,0)
      \egroup
    }
  },
  voltage info/.style = {% 
    append after command = {%
      \bgroup
        [voltage info/.cd,#1]
        [current point is local=true]
        [every voltage info/.try]
        [shift={($(\tikzlastnode.north)+(0,\pgfkeysvalueof{/tikz/voltage info/vertical offset}\tikzcircuitssizeunit)$)}]
        [voltage arrow,\pgfkeysvalueof{/tikz/voltage info/arrow direction}]
        (-\pgfkeysvalueof{/tikz/voltage info/horizontal offset}\tikzcircuitssizeunit,0)
        edge[line to] 
        node[\pgfkeysvalueof{/tikz/voltage info/label
          position},auto,rotate = \pgfkeysvalueof{/tikz/voltage
          info/label rotate}] {\pgfkeysvalueof{/tikz/voltage info/label}}
        (\pgfkeysvalueof{/tikz/voltage info/horizontal offset}\tikzcircuitssizeunit,0)
      \egroup
    }
  },
  voltage info'/.style = {% 
    append after command = {%
      \bgroup
        [voltage info/.cd,label position = below,#1]
        [current point is local=true]
        [every voltage info/.try]
        [shift={($(\tikzlastnode.south)+(0,-\pgfkeysvalueof{/tikz/voltage info/vertical offset}\tikzcircuitssizeunit)$)}]
        [voltage arrow,\pgfkeysvalueof{/tikz/voltage info/arrow direction}]
        (-\pgfkeysvalueof{/tikz/voltage info/horizontal offset}\tikzcircuitssizeunit,0)
        edge[line to] 
        node[\pgfkeysvalueof{/tikz/voltage info/label
          position},auto,swap,rotate = \pgfkeysvalueof{/tikz/voltage
          info/label rotate}] {\pgfkeysvalueof{/tikz/voltage info/label}}
        (\pgfkeysvalueof{/tikz/voltage info/horizontal offset}\tikzcircuitssizeunit,0)
      \egroup
    }
  }
}

\tikzset{circuit declare symbol=source,
     set source graphic={
        draw,
        circuit symbol lines,
        circuit symbol size=width 3 height 3,
        shape=generic circle IEC,
        transform shape
        },
    AC/.style={
        /pgf/generic circle IEC/before background={
            \pgftransformresetnontranslations
            \pgfpathmoveto{\pgfpoint{-0.75\tikzcircuitssizeunit}{0pt}}
            \pgfpathsine{
                \pgfpoint{0.375\tikzcircuitssizeunit}{0.375\tikzcircuitssizeunit}}
            \pgfpathcosine{
                \pgfpoint{0.375\tikzcircuitssizeunit}{-0.375\tikzcircuitssizeunit}}
            \pgfpathsine{
                \pgfpoint{0.375\tikzcircuitssizeunit}{-0.375\tikzcircuitssizeunit}}
            \pgfpathcosine{
                \pgfpoint{0.375\tikzcircuitssizeunit}{0.375\tikzcircuitssizeunit}}
            \pgfusepathqstroke
        }
    }
}

\tikzset{circuit declare symbol=source,
     set source graphic={
        draw,
        circuit symbol lines,
        circuit symbol size=width 3 height 3,
        shape=generic circle IEC,
        transform shape
        },
    DC/.style={
        /pgf/generic circle IEC/before background={
            \pgftransformresetnontranslations
            \pgfpathmoveto{\pgfpointorigin}%\pgfpoint{-0.75\tikzcircuitssizeunit}{0pt}}
            \pgfpathlineto{\pgfpoint{0\tikzcircuitssizeunit}{1.5\tikzcircuitssizeunit}}
						\pgfpathlineto{\pgfpoint{0\tikzcircuitssizeunit}{-1.5\tikzcircuitssizeunit}}
            \pgfusepathqstroke
        }
    }
}

\tikzstyle{block} = [draw,rectangle,minimum height=1cm,minimum width=1cm] %minimum height=2cm,minimum width=2cm
\tikzstyle{doubleblock} = [draw,rectangle,double,double distance=1pt,minimum height=1cm,minimum width=1cm] %minimum height=2cm,minimum width=2cm,
\tikzstyle{sum} = [draw,circle,inner sep=0mm,minimum size=0.4cm]
\tikzstyle{connector} = [->]	
\tikzstyle{line} = []
\tikzstyle{branch} = [circle,inner sep=0pt,minimum size=1mm,fill=black,draw=black]
\tikzstyle{guide} = []
\tikzstyle{snakeline} = [connector, decorate, decoration={pre length=0.2cm,
                         post length=0.2cm, snake, amplitude=.4mm,
                         segment length=2mm},magenta, ->]
\tikzstyle{input} = [coordinate]
\tikzstyle{output} = [coordinate]

%% file: include/MyEnvironments_Paper_EN.tex
\let\c@theorem\relax

\newtheorem{theorem}{Theorem}
\newtheorem{proposition}{Proposition}
\newtheorem{corollary}{Corollary}
\newtheorem{lemma}{Lemma}
\newtheorem{remark}{Remark}
\newtheorem*{remark*}{Remark}

\newtheorem*{fact}{Fact}

\newtheorem{example}{Example}
\newtheorem*{example*}{Example}
\newtheorem{examples}{Examples}
\newtheorem*{examples*}{Examples}
\newtheorem*{assumption*}{Assumption}
\newtheorem{definition}{Definition}
\newtheorem*{definition*}{Definition}
\newtheorem*{definitions*}{Definitions}
\newtheorem{case}{Case}

%% file: include/MySymbols_GeneralMath.tex
\ifx \chosenlanguage\germanlanguage
\else
\fi

\newcommand{\N}{\mathbb{N}}

\newcommand{\R}{\mathbb{R}}

\newcommand{\cset}[1]{\mathfrak{#1}}
\newcommand{\mv}[1]{#1}
\newcommand{\mm}[1]{#1}
\newcommand{\ddt}{\tfrac{{\rm d}}{{\rm d}t}}
\newcommand{\ddtsmall}{\tfrac{{\rm d}}{{\rm d}t}}

%% file: include/MySymbols_ControlSystems.tex
\newcommand{\myref}{\rm ref}

%% file: include/MySymbols_ThreePhaseSystems.tex
\newcommand{\Tp}{\mm{T}_{\!\rm p}} % Park-Transformation matrix 3x3
\newcommand{\J}{\mm{J}} % Park-Transformation matrix for phi=90°
\newcommand{\phik}{\phi}
\newcommand{\omegak}{\omega}

\newcommand{\usref}{\mv{u}^s_{\myref}}

%% file: include/MySymbols_PowerElectronics.tex
\newcommand{\Ts}{T_{\rm s}}

\renewcommand{\sc}{s^c}

\newcommand{\dc}{{\rm dc}}
\newcommand{\udc}{u_{\dc}}

%% file: include/MySymbols_ElectricalDrives.tex
\newcommand{\nom}{\rm nom}

\newcommand{\omegam}{\omega_{\rm m}}

\newcommand{\omegaMnom}{\omega_{{\rm m},\nom}}

\newcommand{\np}{n_{\rm p}}

\newcommand{\mmotor}{m_{\rm m}}
\newcommand{\mload}{m_{\rm l}}

\newcommand{\mMnom}{m_{{\rm m},\nom}}

\newcommand{\Lsd}{L_{{\rm s}}^{\rm d}} % Inductance PMSM d-component	
\newcommand{\Lsq}{L_{{\rm s}}^{\rm q}} % Inductance PMSM q-component

\newcommand{\Rs}{R_{\rm s}}

\newcommand{\isk}{\mv{i}_{{\rm s}}}

\newcommand{\isd}{i_{{\rm s}}^{\rm d}}
\newcommand{\isq}{i_{{\rm s}}^{\rm q}}

\newcommand{\uss}{\mv{u}_{{\rm s}}^{\rm s}}

\newcommand{\usk}{\mv{u}_{{\rm s}}}
\newcommand{\usd}{u_{{\rm s}}^{\rm d}}
\newcommand{\usq}{u_{{\rm s}}^{\rm q}}
\newcommand{\psisk}{\mv{\psi}_{{\rm s}}}
\newcommand{\Psisk}{\mv{\Psi}_{{\rm s}}}

\newcommand{\psisd}{\psi_{{\rm s}}^{\rm d}}
\newcommand{\psisq}{\psi_{{\rm s}}^{\rm q}}

\newcommand{\ismax}{\hat{\imath}_{{\rm s},\max}}

\newcommand{\usmax}{\hat{u}_{{\rm s},\max}}

\newcommand{\ussref}{\mv{u}_{{\rm s},\myref}^{\rm s}}

\newcommand{\ssabc}{\mv{s}_{\rm s}^{abc}}
\newcommand{\ssa}{s_{\rm s}^{\rm a}}
\newcommand{\ssb}{s_{\rm s}^{\rm b}}
\newcommand{\ssc}{s_{\rm s}^{\rm c}}

\newcommand{\vk}{\mv{v}}
\newcommand{\vd}{v^{\rm d}}
\newcommand{\vq}{v^{\rm q}}

\newcommand{\ThetaVecd}{\mv{\theta}_\mathrm{d}}
\newcommand{\ThetaVecq}{\mv{\theta}_\mathrm{q}}

\newcommand{\isdbar}[1]{\bar{i}_{{\rm s},#1}^{\rm d}}

\newcommand{\isqbar}[1]{\bar{i}_{{\rm s},#1}^{\rm q}}

\newcommand{\Psisd}{\Psi_{{\rm s}}^{\rm d}}
\newcommand{\Psisq}{\Psi_{{\rm s}}^{\rm q}}

\newcommand{\Psisdhat}{\hat{\Psi}_{{\rm s}}^{\rm d}}
\newcommand{\Psisqhat}{\hat{\Psi}_{{\rm s}}^{\rm q}}

%% file: include/MySymbols_RSM_JulianKullick.tex
\newcommand{\LsDQ}{L_{{\rm s}}^{d q }} % Inductance RSM dq-component
\newcommand{\LsQD}{L_{{\rm s}}^{q d }} % Inductance RSM qd-component